\documentclass[aps,prd,twoside,twocolumn,nofootinbib,showpacs,floatfix,superscriptaddress,showkeys]{revtex4-1}
\usepackage{amssymb}
\usepackage{amsmath,amssymb}
\usepackage{graphicx,bm}
\usepackage{slashed}
\usepackage{times}
\usepackage{epstopdf}
\usepackage{ulem} 
\usepackage[usenames]{color}
\usepackage{float}
\usepackage{subfigure}
\usepackage{subfigure}
\usepackage{rotating}
\usepackage{color}
\usepackage{multirow}
\usepackage{dcolumn}
\usepackage{overpic}
\usepackage{booktabs}
\usepackage{makecell}
\usepackage{diagbox}
\allowdisplaybreaks[4]

\renewcommand\sout{\bgroup \color{red} \ULdepth=-.5ex \ULset}

\usepackage[colorlinks, citecolor=blue,anchorcolor=red,menucolor=red,linkcolor=blue,filecolor=red,runcolor=red,urlcolor=blue,frenchlinks=red]{hyperref}
\begin{document}
\title{Heavy flavor molecular states with strangeness}

\author{Kan Chen}
\affiliation{School of Physics and Center of High Energy Physics, Peking University, Beijing 100871, China}

\author{Bo Wang}\email{Corresponding author: wangbo@hbu.edu.cn}
\affiliation{School of Physical Science and Technology, Hebei University, Baoding 071002, China}
\affiliation{Key Laboratory of High-precision Computation and Application of Quantum Field Theory of Hebei Province, Baoding 071002, China}

\author{Shi-Lin Zhu}\email{Corresponding author: zhusl@pku.edu.cn}
\affiliation{School of Physics and Center of High Energy Physics, Peking University, Beijing 100871, China}

\begin{abstract}
We proposed a unified framework to describe the interactions of the
observed $T_{cc}$, $P_c$, and $P_{cs}$ within a quark level
interaction in our previous work. In this work, we generalize our
framework to the loosely bound hadronic molecules composed of heavy
flavor di-hadrons with strangeness. We predict the possible
$D^{(*)}D^{(*)}_s$ molecular states in the SU(3) limit with the
masses of the $P_c$ states as the inputs. We also investigate the
baryon-meson and baryon-baryon systems and consider the SU(3)
breaking effect in their flavor wave functions. We generalize our
isospin criterion of the formation of heavy flavor di-hadron
molecules to the $U/V$ spin case. For a specific heavy flavor
meson-meson, baryon-meson, or baryon-baryon system, the interactions
for the states with the same flavor and spin matrix elements can be
related by a generalized flavor-spin symmetry.
\end{abstract}

\maketitle

\vspace{2cm}

\section{Introduction}\label{sec1}

The study of exotic hadrons with configurations beyond the conventional
quark model has become a hot topic since the discovery of $X(3872)$
\cite{Belle:2003nnu}. Especially, the $P_c$
\cite{LHCb:2015yax,LHCb:2019kea}, $P_{cs}$ \cite{LHCb:2020jpq}
 together with the recently reported $T_{cc}$
\cite{LHCb:2021auc,LHCb:2021vvq} are manifestly exotic states.
The observation of these states indicates that the interactions
between heavy flavor di-hadrons share very similar binding
mechanisms.

Since the observed $P_c$, $P_{cs}$, and $T_{cc}$ are below the
thresholds of the corresponding di-hadron systems from several to
several tens MeVs, a natural explanation is that these states are
molecules. To describe the possible similarities of the binding
mechanisms between the above heavy flavor di-hadron systems, we
proposed a hydrogen-like picture and studied the heavy flavor
meson-meson ($M$--$M$), baryon-meson ($B$--$M$), and baryon-baryon ($B$--$B$)
molecular states in Ref. \cite{Chen:2021cfl}. We assume that the
interactions of heavy flavor di-hadron systems are mainly from the
exchange of the light flavor mesons, which is closely related to the
flavor and spin structures of light degrees of freedom (d.o.f) in the
di-hadron systems.

To address the above molecule picture, we adopt a quark-level
Lagrangian
\cite{Meng:2019nzy,Wang:2019nvm,Wang:2020dhf,Chen:2021cfl} to
describe the interactions of different heavy flavor di-hadron
systems. Then the corresponding effective potentials can be
parameterized in terms of the light quark-quark coupling constants
and the flavor and spin matrix elements which are related to its
light d.o.f. In this framework, we obtain a satisfactory description
of the $T_{cc}$, $P_c$, $P_{cs}$ states, and other possible bound
states in the heavy flavor meson-meson, baryon-meson, and
baryon-baryon di-hadron systems which arise from the isospin singlet
and triplet meson exchange interactions. The interactions of
different heavy flavor di-hadron systems ($M$--$M$, $B$--$M$, and $B$--$B$)
can be understood in a molecule picture simultaneously.

\begin{table}[!htbp]
\renewcommand\arraystretch{1.5}
\caption{The heavy flavor meson-meson ($M$--$M$), baryon-meson ($B$--$M$), and baryon-baryon ($B$--$B$) systems ($[H_1H_2]^{I}_{J}$) considered in Ref. \cite{Chen:2021cfl} (with boldface) and those will be considered in this work (with normal font).\label{systems}} 
{
\begin{tabular}{ccccccccccccccccccc}
\toprule[0.8pt]
\multirow{2}{*}{$M$--$M$}&$\boldsymbol{DD}$&$\boldsymbol{DD^*}$&$\boldsymbol{D^*D^*}$ &$DD_s$&$DD^*_s$ ($D_sD^*$)&$D^*D_s^*$\\
&$D_sD_s$&$D_sD_s^*$&$D_s^*D_s^*$\\
\hline
\multirow{5}{*}{$B$--$M$}&$\boldsymbol{\Lambda_c\bar{D}}$&$\boldsymbol{\Lambda_c\bar{D}^*}$&$\boldsymbol{\Sigma_c\bar{D}}$&$\boldsymbol{\Sigma_c\bar{D}^*}$&$\boldsymbol{\Sigma_c^*\bar{D}}$&$\boldsymbol{\Sigma_c^*\bar{D}^*}$\\
&$\Lambda_c\bar{D}_s$&$\Lambda_c\bar{D}_s^*$&$\Sigma_c\bar{D}_s$&$\Sigma_c\bar{D}^*_s$&$\Sigma_c^*\bar{D}_s$&$\Sigma_c^*\bar{D}_s^*$\\
&$\boldsymbol{\Xi_c\bar{D}}$&$\boldsymbol{\Xi_c\bar{D}^*}$&$\boldsymbol{\Xi_c^\prime \bar{D}}$&$\boldsymbol{\Xi_c^\prime \bar{D}^*}$&$\boldsymbol{\Xi_c^*\bar{D}}$&$\boldsymbol{\Xi_c^*\bar{D}^*}$\\
&$\Xi_c\bar{D}_s$&$\Xi_c\bar{D}_s^*$&$\Xi_c^\prime \bar{D}_s$&$\Xi^\prime_c\bar{D}_s^*$&$\Xi_c^*\bar{D}_s$&$\Xi_c^*\bar{D}_s^*$\\
&$\Omega_c\bar{D}$&$\Omega_c\bar{D}^*$&$\Omega_c\bar{D}_s$&$\Omega_c\bar{D}^*_s$\\
\hline
\multirow{5}{*}{$B$--$B$}&$\boldsymbol{\Lambda_c\Lambda_c}$&$\boldsymbol{\Lambda_c\Sigma_c}$&$\Lambda_c\Sigma_c^*$
&$\boldsymbol{\Sigma_c\Sigma_c}$&$\boldsymbol{\Sigma_c\Sigma_c^*}$&$\boldsymbol{\Sigma_c^*\Sigma_c^*}$\\
&$\boldsymbol{\Xi_c\Xi_c}$&$\boldsymbol{\Xi_c\Xi_c^\prime}$&$\boldsymbol{\Xi_c\Xi_c^*}$
&$\boldsymbol{\Xi_c^\prime\Xi_c^\prime}$&$\boldsymbol{\Xi_c^\prime\Xi_c^*}$&$\boldsymbol{\Xi_c^*\Xi_c^*}$\\
&$\Lambda_c\Xi_c$&$\Lambda_c\Xi_c^*$&$\Lambda_c\Xi_c^*$&$\Lambda_c\Omega_c$&$\Sigma_c\Xi_c$&$\Sigma_c\Xi_c^*$\\
&$\Sigma_c\Xi_c^*$&$\Sigma_c\Omega_c$&$\Sigma_c^*\Xi_c$&$\Sigma_c^*\Xi_c^\prime$&$\Sigma_c^*\Xi_c^*$&$\Sigma_c^*\Omega_c$\\
&$\Xi_c\Omega_c$&$\Xi_c^\prime\Omega_c$&$\Xi_c^*\Omega_c$&$\Xi_c^*\Omega_c$&$\Omega_c\Omega_c$\\
\bottomrule[0.8pt]
\end{tabular}}
\end{table}

The simple Lagrangian introduced in our work possesses two important
advantages in Ref. \cite{Chen:2021cfl}. It not only relates the
interactions of different heavy flavor di-hadron ($M$--$M$, $B$--$M$, and
$B$--$B$) systems via their flavor-spin structures but also relates the
interactions of the multiplets in a specific heavy flavor di-hadron
system. In Table \ref{systems}, we list the systems which
were already studied in Ref. \cite{Chen:2021cfl}. In this work, we
further adopt this framework to study the rest of the doubly heavy
flavor di-hadron systems. The
inclusion of the rest of the doubly heavy flavor di-hadron systems
allows us to present a full analysis of the possible flavor-spin
symmetry and the interactions of molecular multiplets in a specific
di-hadron systems. This is the main task of this work.

Some of the doubly heavy tetraquark states with strangeness have
already been discussed in literature. For example, the strange
doubly heavy tetraquark states in molecular configurations have been
discussed in Refs.
\cite{Dong:2021bvy,Molina:2010tx,Dai:2021vgf,Sun:2012sy,SanchezSanchez:2017xtl,Ding:2021igr,Deng:2021gnb}.
The compact strange tetraquark states were studied via various
methods, such as quark potential models
\cite{Meng:2020knc,Lu:2020rog,Deng:2018kly,Anwar:2018sol}, lattice
QCD \cite{Bicudo:2015vta}, chromo-magnetic interaction model (CMI)
\cite{Luo:2017eub,Guo:2021yws,Cheng:2020nho,Eichten:2017ffp}, and
QCD sum rule \cite{Agaev:2019lwh,Agaev:2018vag}.

The strange doubly heavy baryon-meson and di-baryon systems have also been discussed in
literature. The possible pentaquark states composed of
$qqqc\bar{c}$ configurations with $q=u$, $d$, and $s$ were studied
within the framework of CMI model \cite{Wu:2017weo}. The OBE model
was adopted to study the interactions of the $\Xi_c^{(\prime,
*)}\bar{D}_s^{(*)}$ \cite{Wang:2020bjt} and
$\Omega_c^{(*)}\bar{D}_s^{(*)}$ \cite{Wang:2021hql} systems. In Ref.
\cite{Shi:2021wyt}, the hidden charm pentaquark states composed of
$qqsc\bar{c}$, $qssc\bar{c}$, and $sssc\bar{c}$ configurations are
studied in a diquark-diquark-antiquark model. A general discussion
of the heavy-antiheavy hadronic molecules was performed in Ref.
\cite{Dong:2021juy}. In addition, the authors in Ref.
\cite{Li:2012bt} adopted the OBE model to discuss the di-baryon
systems composed of spin-$\frac{1}{2}/\frac{3}{2}$ and
flavor-$\bm{\bar{3}}/\bm{6}$ single heavy baryons systematically.
A general discussion on these systems based on the resonance
saturation model can be found in Ref. \cite{Dong:2021bvy}.

This paper is organized as follows. In Sec. \ref{sec2}, we introduce
our theoretical framework. Then we present our numerical results and
discussions in Sec. \ref{sec3}. In Sec. \ref{sec4}, we conclude this
work with a short summary.

\section{Theoretical framework}\label{sec2}

The interactions for the heavy flavor di-hadron systems listed in
Table \ref{systems} can be calculated by considering the possible
contributions from the exchanges of light mesons, which depend on
the flavor and spin structures of the heavy flavor di-hadron systems
from their light d.o.f. On the other hand, the $S$-wave interactions
of the ground heavy flavor di-hadron systems may arise from the
exchange of the $J^P=0^+$ scalar or $1^+$ axial vector mesons. Thus,
we introduce the following quark level Lagrangians
\cite{Meng:2019nzy,Wang:2019nvm,Wang:2020dhf,Chen:2021cfl}
\begin{eqnarray}
\mathcal{L}=g_s\bar{q}\mathcal{S}q+g_a\bar{q}\gamma_\mu\gamma^5\mathcal{A}^\mu
q \label{Lagrangian}
\end{eqnarray}
to describe the interactions among the considered ground heavy
flavor di-hadron systems. Here, $q=(u, d, s)$, $g_s$ and $g_a$ are
two independent coupling constants that encode the nonperturbative
low energy dynamics of the di-hadron systems.

From Eq. (\ref{Lagrangian}), the effective potential of the light
quark-quark interactions reads
\begin{eqnarray}
V_{qq}&=&\tilde{g}_s\bm{\lambda}_1\cdot\bm{\lambda}_2+\tilde{g}_a\bm{\lambda}_1\cdot\bm{\lambda}_2\bm{\sigma}_1\cdot\bm{\sigma}_2\nonumber\\&
=&\tilde{g}_s\left(\lambda_1^8\lambda_2^8+\lambda_1^i\lambda_2^i+\lambda_1^j\lambda_2^j\right)\nonumber\\
&&+\tilde{g}_a\left(\lambda_1^8\lambda_2^8+\lambda_1^i\lambda_2^i+\lambda_1^j\lambda_2^j\right)\bm{\sigma}_1\cdot\bm{\sigma}_2.\label{Vqq}
\end{eqnarray}
Here, $i$ and $j$ sum from 1 to 3 and 4 to 7, respectively. The
$\lambda_1^8\lambda_2^8$, $\lambda_1^i\lambda_2^i$, and
$\lambda_1^j\lambda_2^j$ are the SU(3) flavor matrices of the
isospin singlet, triplet, and two doublets, respectively. The
$\bm{\sigma}_1\cdot\bm{\sigma}_2$ is the Pauli matrix in spin space.
The redefined coupling constants are
$\tilde{g}_s=g_s^2/m_{\mathcal{S}}^2$ and
$\tilde{g}_a=g_a^2/m^2_{\mathcal{A}}$. The local form of the
effective potential $V_{qq}$ is the result of integrating out the
exchanged spurions.

The effective potential in Eq. (\ref{Vqq}) consists of three sets of
flavor related operators and three sets of flavor-spin related
operators. For convenience, we adopt the following notations
\begin{eqnarray}
\mathcal{O}_1&=&\lambda_1^8\lambda_2^8,\quad \mathcal{O}_4=\lambda_1^8\lambda_2^8\bm{\sigma}_1\cdot\bm{\sigma}_2,\nonumber\\
\mathcal{O}_2&=&\lambda_1^i\lambda_2^i,\quad \mathcal{O}_5=\lambda_1^i\lambda_2^i\bm{\sigma}_1\cdot\bm{\sigma}_2,\nonumber\\
\mathcal{O}_3&=&\lambda_1^j\lambda_2^j,\quad
\mathcal{O}_6=\lambda_1^j\lambda_2^j\bm{\sigma}_1\cdot\bm{\sigma}_2.\nonumber
\end{eqnarray}

In our previous work \cite{Chen:2021cfl}, we only consider the
di-hadron systems without the strange meson exchange contribution.
The coupling constants $\tilde{g}_s$ and $\tilde{g}_a$ are obtained
from a solvable Lippmann-Schwinger equation (LSE). For the heavy
flavor di-hadron systems containing strange quark(s), the
contributions from the $\lambda_1^j\lambda_2^j$ ($j=4, 5, 6, 7$)
operator should also be included. Note that in the scalar or
axial-vector SU(3) flavor octet, the physical masses of the
exchanged mesons that belong to the flavor isospin doublets are
close to the masses of their isospin singlet or triplet mesons.
Thus, we take the SU(3) limit and set the coupling parameters
$\tilde{g}_s$ and $\tilde{g}_a$ to be equal to those of the isospin
singlet or triplet mesons.

The effective potential for a specific heavy flavor di-hadron
$H_1H_2$ system with total isospin $I$ and total angular momentum
$J$ can be written as
\begin{eqnarray}
V_{[H_1H_2]_J^I}=\langle[H_1H_2]_J^I|V_{qq}|[H_1H_2]_J^I\rangle,
\label{VHH}
\end{eqnarray}
where $|[H_1H_2]_J^I\rangle$ denotes the quark-level spin-flavor
wave function of the $H_1H_2$ system. Note that as an approximation,
the correction from heavy degree of freedom is omitted. The
operators $\bm{\lambda_1}\cdot\bm{\lambda_2}$ and
$\bm{\sigma}_1\cdot\bm{\sigma}_2$ only act on the flavor and spin
wave functions of the light quark components. As can be seen from
Eq. (\ref{Vqq}), the interactions of the considered di-hadron
systems only depend on the flavor
($\langle\bm{\lambda}_1\cdot\bm{\lambda}_2\rangle$) and spin
($\langle\bm{\sigma}_1\cdot\bm{\sigma}_2\rangle$) matrix elements,
implying a flavor-spin symmetry for the interactions of the heavy
flavor di-hadron systems. In the following, we will further discuss
the properties of this symmetry.

\subsection{The spin matrix elements}

The operator $\bm{\sigma}_1\cdot\bm{\sigma}_2$ can be easily
calculated at hadron level with the technique of angular momentum
theory. At hadron level, the matrix elements of the
$\bm{l}_1\cdot\bm{l}_2$ spin-spin operator in the light d.o.f for
the discussed $M$--$M$, $B$--$M$, and $B$--$B$ systems can be obtained via a
spin rearrangement procedure. The di-hadron systems can be related
to the states with a specific total light spin and a heavy spin by
the relations \cite{Meng:2019ilv}
\begin{eqnarray}
\left|l_1h_1\right.&&\left. S_1l_2h_2S_2JM\right\rangle\nonumber\\
&&=\sum_{L,H}\sqrt{(2S_1+1)(2S_2+1)(2L+1)(2H+1)}\nonumber\\
&&\times\begin{Bmatrix}
           l_1 & l_2 & L \\
           h_1 & h_2 & H \\
           S_1 & S_2 & J \\
         \end{Bmatrix}\left|l_1l_2Lh_1h_2HJM\right\rangle,
\end{eqnarray}
where $l_i$ and $h_i$ are the light and heavy spin for $H_i$ hadron,
respectively. $L$ and $H$ are the total light spin and total heavy
spin of the di-hadron system, respectively. $S_i$ is the total spin
of the $H_i$ hadron. Then the matrix elements of the operator
$\bm{l}_1\cdot\bm{l}_2$ can be obtained as
\begin{eqnarray}
\left\langle\bm{l}_1\cdot\bm{l}_2\right\rangle&=&\sum_{L,H}\left[L\left(L+1\right)-l_1\left(l_1+1\right)-l_2\left(l_2+1\right)\right]\nonumber\\
&&\times\frac{1}{2}\left(2S_1+1\right)\left(2S_2+1\right)\left(2L+1\right)\left(2H+1\right)\nonumber\\
&&\times\begin{Bmatrix}
           l_1 & l_2 & L \\
           h_1 & h_2 & H \\
           S_1 & S_2 & J \\
         \end{Bmatrix}^2
\end{eqnarray}
Since the $h_i$ is identical for all the single heavy flavor
hadrons, we use a more brief notation $|l_1,S_1;l_2,S_2;J\rangle$ to
specify the spin wave function of a $[H_1H_2]_J$ di-hadron system.
In Table \ref{spin ME}, we collect the results of
$\langle\bm{l}_1\cdot\bm{l}_2\rangle$ matrix elements for the
discussed $M$--$M$, $B$--$M$, and $B$--$B$ di-hadron systems.

\begin{table*}[htbp]
\renewcommand\arraystretch{1.5}
\caption{The matrix elements $\langle\bm{l}_1\cdot\bm{l}_2\rangle$
for the discussed $M$--$M$, $B$--$M$, and $B$--$B$ di-hadron systems with
spin wave functions $|l_1,S_1;l_2,S_2;J\rangle$. \label{spin ME}}
\begin{tabular}{ccccccc}
\toprule[0.8pt] 
&Spin&$\langle\bm{l}_1\cdot\bm{l}_2\rangle$&Spin&$\langle\bm{l}_1\cdot\bm{l}_2\rangle$&Spin&$\langle\bm{l}_1\cdot\bm{l}_2\rangle$\\
\hline
\multirow{2}{*}{$M$--$M$}&$\left|\frac{1}{2},0;\frac{1}{2},0;0\right\rangle$&0&$\left|\frac{1}{2},0;\frac{1}{2},1;1\right\rangle$&0&$\left|\frac{1}{2},1;\frac{1}{2},0;1\right\rangle$&0\\
&$\left|\frac{1}{2},1;\frac{1}{2},1;0\right\rangle$&$-2$&$\left|\frac{1}{2},1;\frac{1}{2},1;1\right\rangle$&$-1$&$\left|\frac{1}{2},1;\frac{1}{2},1;2\right\rangle$&1\\
\hline
\multirow{4}{*}{$B$--$M$}&$\left|0,\frac{1}{2};\frac{1}{2},0;\frac{1}{2}\right\rangle$&0&$\left|0,\frac{1}{2};\frac{1}{2},1;\frac{1}{2}\right\rangle$&0&$\left|0,\frac{1}{2};\frac{1}{2},1;\frac{3}{2}\right\rangle$&0\\
&$\left|1,\frac{1}{2};\frac{1}{2},0;\frac{1}{2}\right\rangle$&0&$\left|1,\frac{1}{2};\frac{1}{2},1;\frac{1}{2}\right\rangle$&$-\frac{8}{3}$&$\left|1,\frac{1}{2};\frac{1}{2},1;\frac{3}{2}\right\rangle$&$\frac{4}{3}$\\
&$\left|1,\frac{3}{2};\frac{1}{2},0;\frac{3}{2}\right\rangle$&0&$\left|1,\frac{3}{2};\frac{1}{2},1;\frac{1}{2}\right\rangle$&$-\frac{10}{3}$&$\left|1,\frac{3}{2};\frac{1}{2},1;\frac{3}{2}\right\rangle$&$-\frac{4}{3}$\\
&$\left|1,\frac{3}{2};\frac{1}{2},1;\frac{5}{2}\right\rangle$&2&&&&                                 \\
\hline
\multirow{7}{*}{$B$--$B$}&$\left|0,\frac{1}{2};0,\frac{1}{2};0\right\rangle$&0&$\left|0,\frac{1}{2};0,\frac{1}{2};1\right\rangle$&0&$\left|0,\frac{1}{2};1,\frac{1}{2};0\right\rangle$&0\\
&$\left|0,\frac{1}{2};1,\frac{1}{2};1\right\rangle$&0&$\left|1,\frac{1}{2};0,\frac{1}{2};0\right\rangle$&0&$\left|1,\frac{1}{2};0,\frac{1}{2};1\right\rangle$&0\\
&$\left|0,\frac{1}{2};1,\frac{3}{2};1\right\rangle$&1&$\left|0,\frac{1}{2};1,\frac{3}{2};2\right\rangle$&0&$\left|1,\frac{3}{2};0,\frac{1}{2};1\right\rangle$&0\\
&$\left|1,\frac{3}{2};0,\frac{1}{2};2\right\rangle$&0&$\left|1,\frac{1}{2};1,\frac{1}{2};0\right\rangle$&$-\frac{16}{3}$&$\left|1,\frac{1}{2};1,\frac{1}{2};1\right\rangle$&$\frac{16}{9}$\\
&$\left|1,\frac{1}{2};1,\frac{3}{2};1\right\rangle$&-$\frac{40}{9}$&$\left|1,\frac{1}{2};1,\frac{3}{2};2\right\rangle$&$\frac{8}{3}$&$\left|1,\frac{3}{2};1,\frac{1}{2};1\right\rangle$&$-\frac{40}{9}$\\
&$\left|1,\frac{3}{2};1,\frac{1}{2};2\right\rangle$&$\frac{8}{3}$&$\left|1,\frac{3}{2};1,\frac{3}{2};0\right\rangle$&$-\frac{20}{3}$&$\left|1,\frac{3}{2};1,\frac{3}{2};1\right\rangle$&$-\frac{44}{9}$\\
&$\left|1,\frac{3}{2};1,\frac{3}{2};2\right\rangle$&$-\frac{4}{3}$&$\left|1,\frac{3}{2};1,\frac{3}{2};3\right\rangle$&4&&\\
\bottomrule[0.8pt]
\end{tabular}
\end{table*}

\subsection{The flavor matrix elements in the SU(3) limit}

Now we discuss the quark-level flavor wave functions of the
considered di-hadron systems. In the SU(3) limit, the two
anti-charmed mesons belong to the flavor $\bm{3}$ representation and
have
\begin{eqnarray}
\bm{3}\otimes\bm{3}=\bm{6}\oplus\bar{\bm{3}}. \label{MMF}
\end{eqnarray}
Similarly, a charmed baryon and an anti-charmed meson can form the
following representations
\begin{eqnarray}
\left(\bm{3}\otimes\bm{3}\right)\otimes \bm{3}&=&\left(\bm{6}\otimes\bm{3}\right)\oplus\left(\bar{\bm{3}}\otimes\bm{3}\right)\nonumber\\
&=&\left(\bm{8}\oplus\bm{10}\right)\oplus\left(\bm{1}\oplus\bm{8}^\prime\right).
\label{BMF}
\end{eqnarray}
Two singly charmed baryons have four light quarks. The product of
four $\bm{3}$ representations can be divided into
\begin{eqnarray}
&&\left(\bm{3}\otimes\bm{3}\right)\otimes\left(\bm{3}\otimes\bm{3}\right)\nonumber\\
=&&\left(\bm{6}\otimes\bm{6}\right)\oplus\left(\bm{6}\otimes\bar{\bm{3}}\right)\oplus\left(\bar{\bm{3}}\otimes\bm{6}\right)\oplus\left(\bar{\bm{3}}\otimes\bar{\bm{3}}\right)\nonumber\\
=&&\left(\bm{15}_1\oplus\bm{15}_2\oplus\bar{\bm{6}}_1\right)\oplus\left(\bm{15}\oplus\bm{3}_1\right)\oplus\left(\bm{15}\oplus\bm{3}_1\right)\oplus\left(\bm{3}_2\oplus\bar{\bm{6}}_2\right).\nonumber\\
\label{BBF}
\end{eqnarray}
We construct the explicit forms of flavor wave functions for the
multiplets of different heavy flavor di-hadron systems in the Eqs.
(\ref{MMF})-(\ref{BBF}) using the SU(3) Clebsch-Gordan (CG) coefficients
\cite{Kaeding:1995vq}. We collect the results of the
$\langle\bm{\lambda}_1\cdot\bm{\lambda}_2\rangle$ matrix elements
for the different di-hadron systems in their corresponding SU(3)
representations in Table \ref{flavor ME}. For the di-hadron systems
in the same SU(3) multiplet, their
$\langle\bm{\lambda}_1\cdot\bm{\lambda}_2\rangle$ matrix elements
have the same value.

\begin{table}[htbp]
\renewcommand\arraystretch{1.5}
\caption{The matrix elements
$\langle\bm{\lambda}_1\cdot\bm{\lambda}_2 \rangle$ for the discussed
$M$--$M$, $B$--$M$, and $B$--$B$ di-hadron systems in different SU(3)
multiplets. \label{flavor ME}}
\begin{tabular}{cccccccc}
\toprule[0.8pt]
$M$--$M$&$\bm{6}$&$\bar{\bm{3}}$&&&&&\\
$\left\langle\bm{\lambda}_1\cdot\bm{\lambda}_2\right\rangle$&$\frac{4}{3}$&$-\frac{8}{3}$&&&&&\\
\hline
$B$--$M$&$\bm{8}$&$\bm{10}$&$\bm{1}$&$\bm{8}^\prime$&&&\\
$\left\langle\bm{\lambda}_1\cdot\bm{\lambda}_2\right\rangle$&$-\frac{10}{3}$&$\frac{8}{3}$&$-\frac{16}{3}$&$\frac{2}{3}$&&&\\
\hline
$B$--$B$&$\bm{15}_1$&$\bm{15}_2$&$\bar{\bm{6}}_1$&$\bm{15}$&$\bm{3}_1$&$\bm{3}_2$&$\bar{\bm{6}}_2$\\
$\left\langle\bm{\lambda}_1\cdot\bm{\lambda}_2\right\rangle$&$-\frac{8}{3}$&$\frac{16}{3}$&$-\frac{20}{3}$&$\frac{4}{3}$&$-\frac{20}{3}$&$-\frac{8}{3}$&$\frac{4}{3}$\\
\bottomrule[0.8pt]
\end{tabular}
\end{table}

Combining the results in Table \ref{spin ME} and Table \ref{flavor
ME}, we emphasize that the flavor-spin symmetry in Eq. (\ref{Vqq})
demonstrates that if two heavy flavor di-hadron systems have the
same spin structure $|l_1,S_1;l_2,S_2;J\rangle$ and belong to the
same SU(3) multiplet, they have the same effective potentials.
Correspondingly, if the potential can provide enough attractions to
form a bound state, their binding energies should have the same size.

In addition, we can extract many other useful information
numerically from Table \ref{spin ME} and Table \ref{flavor ME}. For
example, as presented in Table \ref{flavor ME}, in the $B$--$B$
systems, the matrix elements
$\langle\bm{\lambda}_1\cdot\bm{\lambda}_2\rangle$ for the $\bm{15}$
and $\bar{\bm{6}}_2$ multiplets are the same. Thus, if their spin
matrix elements turn out to be the same (it is easy to find two
different spin states that have the same $\langle
\bm{l}_1\cdot\bm{l}_2\rangle$ in Table \ref{spin ME}), they should
have the same effective potentials. Thus, the systems which share
the same values of the
$\langle\bm{\lambda}_1\cdot\bm{\lambda}_2\rangle$ and
$\langle\bm{\sigma}_1\cdot\bm{\sigma}_2\rangle$ matrix elements
should have the same effective potential and are related by this
generalized flavor-spin symmetry.

\subsection{The wave functions of heavy flavor di-hadron systems including the SU(3) breaking effect}

In the previous section, we assume a perfect SU(3) symmetry when we
construct the flavor wave functions of the heavy flavor di-hadron
systems. To discuss the physical di-hadron systems, we need to
distinguish the $s$ quark from $u$, $d$ quarks when we construct the
corresponding flavor wave functions. In Table \ref{flavor WF}, we
collect the quark-level flavor wave functions of the charmed hadrons
considered in this work.
\begin{table}[htbp]
\setlength\tabcolsep{0.9pt} \caption{The flavor wave functions for
the charmed hadrons considered in this work. \label{flavor WF}}
\begin{tabular}{llcllc}
\toprule[0.8pt]
Hadron&$\left|II_z\right\rangle$&$\phi^{H_f}_{Im_I}$&Hadron&$\left|II_z\right\rangle$&$\phi^{H_f}_{Im_I}$\\
\hline
$\bar{D}^{(*)0}$&$\left|\frac{1}{2}\frac{1}{2}\right\rangle$&$u\bar{c}$&$\bar{D}^{(*)-}$&$\left|\frac{1}{2}-\frac{1}{2}\right\rangle$&$d\bar{c}$\\
$\bar{D}_s^{(*)-}$&$\left|00\right\rangle$&$s\bar{c}$&$\Lambda_c^+$&$\left|00\right\rangle$&$\frac{1}{\sqrt{2}}\left(du-ud\right)c$\\
$\Sigma_c^{(*)++}$&$\left|11\right\rangle$&$\frac{1}{\sqrt{2}}uuc$&$\Sigma_c^{(*)+}$&$\left|10\right\rangle$&$\frac{1}{\sqrt{2}}\left(ud+du\right)c$\\
$\Sigma_c^{(*)0}$&$\left|1-1\right\rangle$&$ddc$&$\Xi_c^+$&$\left|\frac{1}{2}\frac{1}{2}\right\rangle$&$\frac{1}{\sqrt{2}}\left(us-su\right)c$\\
$\Xi_c^0$&$\left|\frac{1}{2}-\frac{1}{2}\right\rangle$&$\frac{1}{\sqrt{2}}\left(ds-sd\right)c$&$\Xi_c^{\prime(*)+}$&$\left|\frac{1}{2}\frac{1}{2}\right\rangle$&$\frac{1}{\sqrt{2}}\left(us+su\right)c$\\
$\Xi_c^{\prime(*)0}$&$\left|\frac{1}{2}-\frac{1}{2}\right\rangle$&$\frac{1}{\sqrt{2}}\left(ds+sd\right)c$&$\Omega_c^0$&$\left|00\right\rangle$&$ssc$\\
\bottomrule[0.8pt]
\end{tabular}
\end{table}

Then we construct the total wave functions of the considered
heavy-flavor baryon-meson systems as
\begin{eqnarray}
|[H_1H_2]_J^I\rangle&=&\sum_{m_{I_1},m_{I_2}}C_{I_1,m_{I_1};I_2,m_{I_2}}^{I,I_z}\phi_{I_1,m_{I_1}}^{H_{1f}}\phi_{I_2,m_{I_2}}^{H_{2f}}\nonumber\\
&&\otimes\sum_{m_{S_1},m_{S_2}}C^{J,J_z}_{S_1,m_{S_1};S_2,m_{S_2}}\phi_{S_1,m_{S_1}}^{H_{1s}}\phi_{S_2,m_{S_2}}^{H_{2s}}.\nonumber\\
\label{BMFSU2}
\end{eqnarray}
Here, the spin wave functions $\phi_{S_1,m_{S_1}}^{H_{1s}}$ and
$\phi_{S_2,m_{S_2}}^{H_{2s}}$ can be constructed directly using the
SU(2) CG coefficients. For the flavor wave functions
$\phi_{I_1,m_{I_1}}^{H_{1f}}$ and $\phi_{I_2,m_{I_2}}^{H_{2f}}$, we
adopt the flavor wave functions listed in Table \ref{flavor WF} and
use the SU(2) CG coefficients to construct the total flavor wave
functions of considered $B$--$M$ systems. Unlike using the SU(3) CG
coefficients, the construction in Eq. (\ref{BMFSU2}) ensures that
the $s$ quark can not interchange with the $u$ or $d$ quark and acts
as a SU(3) flavor singlet.

For the flavor wave functions of the $M$--$M$ and $B$--$B$ systems, we
first construct the quark-level $|H_1H_2\rangle^{I}$ and
$|H_2H_1\rangle^I$ flavor wave functions with SU(2) CG coefficients.
Then we combine them with a (an) symmetric (anti-symmetric) factor
to partly include the symmetric properties introduced from the $s$
quark. Collectively, the flavor-spin wave functions for the heavy
flavor meson-meson and baryon-baryon systems are
\begin{eqnarray}
|[H_1H_2]_J^{IS/A}\rangle&=&\sum_{m_{I_1},m_{I_2}}\frac{1}{\sqrt{2}}\left(C_{I_1,m_{I_1};I_2m_{I_2}}^{I,I_z}\phi_{I_1,m_{I_1}}^{H_{1f}}\phi_{I_2,m_{I_2}}^{H_{2f}}\right.\nonumber\\&&\left.
\pm C_{I_2,m_{I_2};I_1m_{I_1}}^{I,I_z}\phi_{I_2,m_{I_2}}^{H_{2f}}\phi_{I_1,m_{I_1}}^{H_{1f}}\right)\nonumber\\&&\otimes\sum_{m_{S_1},m_{S_2}}C_{S_1,m_{S_1};S_2,m_{S_2}}^{J,J_z}\phi_{S_1,m_{S_1}}^{H_{1s}}\phi_{S_2,m_{S_2}}^{H_{2s}}.\nonumber\\
\label{MMBBFSU2}
\end{eqnarray}
where the $1/\sqrt{2}$ is the normalization factor, and the
superscript $S$ ($A$) corresponds to the symmetric factor $+$ ($-$).

After considering the SU(3) breaking effect, some of the flavor
functions constructed from Eqs. (\ref{BMFSU2})-(\ref{MMBBFSU2}) can
still correspond to the flavor wave functions constructed in the
SU(3) limit. For example, the flavor wave function for the
$\Xi_c\bar{D}$ isospin triplet constructed from Eq. (\ref{BMFSU2})
is identical to the SU(3) $|RYI\rangle=|\bm{8}^\prime 0 1\rangle$
triplet in the $B$--$M$ sector, where $R$, $Y$, and $I$ denote the
dimension of the SU(3) representation, hypercharge, and isospin,
respectively. They have the same
$\langle\bm{\lambda}_1\cdot\bm{\lambda}_2\rangle$ matrix elements.
On the contrary, in most cases, the flavor functions constructed
from Eqs. (\ref{BMFSU2})-(\ref{MMBBFSU2}) do not have corresponding
SU(3) symmetry states. We need to further calculate their
$\langle\bm{\lambda}_1\cdot\bm{\lambda}_2\rangle$ matrix elements.
We will present our results in the next section.

After we deduce the effective potentials for all the studied heavy
flavor di-hadron systems, we iterate the corresponding effective
potentials into the Lippmann-Schwinger equation (LSE),
\begin{eqnarray}
T\left(p^\prime,p\right)=V(p^\prime,p)+\int\frac{d^3q}{(2\pi)^3}\frac{V\left(p^\prime,q\right)T\left(q,p\right)}{E-\frac{q^2}{2m_\mu}+i\epsilon},
\end{eqnarray}
where $m_\mu$ is the reduced mass of the di-hadron system. $p$ and
$p^\prime$ are the momentum of the initial and final states in the
center of mass frame, respectively.

We introduce a hard regulator to exclude the contribution from
higher momenta \cite{Meng:2021jnw,Meng:2021kmi},
\begin{eqnarray}
V\left(p^\prime,p\right)=V_{[H_1H_2]_J^I}\Theta\left(\Lambda-p\right)\Theta\left(\Lambda-p^\prime\right).
\end{eqnarray}
After some deductions, we can solve the following equation to find
the possible bound state of the $[H_1H_2]_J^I$ system,
\begin{eqnarray}
1-\frac{m_\mu}{\pi^2}V_{\left[H_1H_2\right]^I_J}\left[-\Lambda+k\text{tan}^{-1}\left(\frac{\Lambda}{k}\right)\right]=0,\label{LSE}
\end{eqnarray}
with $k=\sqrt{-2m_\mu E}$.

\section{Numerical results and discussion}\label{sec3}

In our model, we have three free parameters: the quark coupling
constants $\tilde{g}_s$, $\tilde{g}_a$, and the momentum cutoff
parameter $\Lambda$. In our previous work, we use the masses of the
$P_{c}(4312)$, $P_{c}(4440)$, and $P_c(4457)$ to construct an
equation set and precisely solve the above parameters with
$\tilde{g}_s=11.739$ GeV$^{-2}$, $\tilde{g}_a=-2.860$ GeV$^{-2}$,
and $\Lambda=0.409$ GeV. With these parameters, we reproduce the
masses of the recently observed $P_{cs}$ and $T_{cc}$ states very
well. One can refer to Ref. \cite{Chen:2021cfl} for more details. In
our convention, a positive (negative) $V_{[H_1H_2]_J^I}$ corresponds
to a(n) repulsive (attractive) force. In this section, we present
our numerical results of the meson-meson, baryon-meson, and
baryon-baryon systems case by case.

\subsection{Meson-Meson system}

We take the $[\bar{D}\bar{D}_s^*]_1^{\frac{1}{2}S/A}$ (or
equivalently $[\bar{D}_s\bar{D}^*]_1^{\frac{1}{2}S/A}$) states as
examples to show more details of our calculations. From Eq.
(\ref{MMBBFSU2}), the quark-level flavor-spin wave functions for the
$[\bar{D}\bar{D}_s^*]_1^{\frac{1}{2}S/A}$ systems can be directly
written as
\begin{eqnarray}
|[\bar{D}\bar{D}_s^*]_1^{\frac{1}{2}S}\rangle&=&\frac{1}{\sqrt{2}}\left(u\bar{c}s\bar{c}+s\bar{c}u\bar{c}\right)\otimes\frac{1}{\sqrt{2}}\left(\uparrow\downarrow\uparrow\uparrow-\downarrow\uparrow\uparrow\uparrow\right),\nonumber\\
\label{DDSW1}
\\
|[\bar{D}\bar{D}_s^*]_1^{\frac{1}{2}A}\rangle&=&\frac{1}{\sqrt{2}}\left(u\bar{c}s\bar{c}-s\bar{c}u\bar{c}\right)\otimes\frac{1}{\sqrt{2}}\left(\uparrow\downarrow\uparrow\uparrow-\downarrow\uparrow\uparrow\uparrow\right).\nonumber\\
\label{DDSW2}
\end{eqnarray}
Here, we only present their $I_z=1/2$ and $J_z=1$ components. The
other components with the same $I$ and $J$ will give the identical
matrix elements. Then we substitute Eqs. (\ref{DDSW1})-(\ref{DDSW2})
into Eq. (\ref{VHH}), we will encounter the direct terms
\begin{eqnarray}
\left\langle
u\bar{c}s\bar{c}\right|\bm{\lambda}_1\cdot\bm{\lambda}_2\left|u\bar{c}s\bar{c}\right\rangle,\quad
\left\langle
s\bar{c}u\bar{c}\right|\bm{\lambda}_1\cdot\bm{\lambda}_2\left|s\bar{c}u\bar{c}\right\rangle
\end{eqnarray}
that only have non-vanishing matrix elements from the
$\lambda_1^8\lambda_2^8$ term and the cross terms
\begin{eqnarray}
\left\langle
u\bar{c}s\bar{c}\right|\bm{\lambda}_1\cdot\bm{\lambda}_2\left|s\bar{c}u\bar{c}\right\rangle,\quad
\left\langle
s\bar{c}u\bar{c}\right|\bm{\lambda}_1\cdot\bm{\lambda}_2\left|u\bar{c}s\bar{c}\right\rangle
\end{eqnarray}
that only receive contributions from the $\lambda_1^j\lambda_2^j$
components. Thus, for the
$[\bar{D}\bar{D}_s^*]_{1}^{\frac{1}{2}S/A}$ systems, they do not
have interactions from the exchange of the isospin triplet
($\lambda_1^i\lambda_2^i$) fields.

Similarly, the isospin triplet related operators
$\mathcal{O}_2(\lambda_1^i\lambda_2^i)$ and
$\mathcal{O}_5(\lambda_1^i\lambda_2^i\bm{\sigma}_1\cdot\bm{\sigma}_2)$
do not contribute to the interactions of the $\bar{D}\bar{D}_s$ and
$\bar{D}^{*}\bar{D}_s^{*}$ systems. Besides, the interactions of the
$\bar{D}^{(*)}_s\bar{D}_s^{(*)}$ systems only arise from the
$\mathcal{O}_1$ and $\mathcal{O}_4$ operators. In Table \ref{MMME},
we present the matrix elements of the $\mathcal{O}_1$,
$\mathcal{O}_3$, $\mathcal{O}_4$, $\mathcal{O}_6$ for the
$\bar{D}\bar{D}_s$, $\bar{D}\bar{D}^*_s$ ($D_sD^*$) and
$\bar{D}_s^{(*)}\bar{D}_s^{(*)}$ di-meson systems.

\begin{table}[htbp]
\renewcommand\arraystretch{1.5}
\caption{The matrix elements of the operators $\mathcal{O}_1$
($\lambda^8_1\lambda^8_2$), $\mathcal{O}_3$
($\lambda^j_1\lambda^j_2$), $\mathcal{O}_4$
($\lambda^8_1\lambda^8_2\bm{\sigma}_1\cdot\bm{\sigma}_2$), and
$\mathcal{O}_6$
($\lambda^j_1\lambda^j_2\bm{\sigma}_1\cdot\bm{\sigma}_2$) for the
considered heavy flavor di-meson systems listed in Table
\ref{systems}. \label{MMME}}
\begin{tabular}{ccccc}
\toprule[0.8pt]
System&$\mathcal{O}_1$&$\mathcal{O}_3$&$\mathcal{O}_4$&$\mathcal{O}_6$\\
\hline
$[\bar{D}\bar{D}_s]^{\frac{1}{2}S}_{0}$&$-\frac{2}{3}$&2&0&0\\
$[\bar{D}\bar{D}_s]^{\frac{1}{2}A}_{0}$&$-\frac{2}{3}$&-2&0&0\\
$[\bar{D}\bar{D}_s^*(\bar{D}_s\bar{D}^*)]_1^{\frac{1}{2}S}$&$-\frac{2}{3}$&2&0&0\\
$[\bar{D}\bar{D}_s^*(\bar{D}_s\bar{D}^*)]_1^{\frac{1}{2}A}$&$-\frac{2}{3}$&-2&0&0\\
$[\bar{D}^*\bar{D}_s^*]_{0,1,2}^{\frac{1}{2}S}$&$-\frac{2}{3}$/$-\frac{2}{3}$/$-\frac{2}{3}$&2/2/2&$\frac{4}{3}$/$\frac{2}{3}$/$-\frac{2}{3}$&-4/-2/2\\
$[\bar{D}^*\bar{D}_{s}^*]_{0,1,2}^{\frac{1}{2}A}$&$-\frac{2}{3}$/$-\frac{2}{3}$/$-\frac{2}{3}$&-2/-2/-2&$\frac{4}{3}$/$\frac{2}{3}$/$-\frac{2}{3}$&4/2/-2\\
$[\bar{D}_s\bar{D}_s]^0_{0}$&$\frac{4}{3}$&0&0&0\\
$[\bar{D}_s\bar{D}_s^*]_1^0$&$\frac{4}{3}$&0&0&0\\
$[\bar{D}^*_s\bar{D}_s^*]^0_{0,2}$&$\frac{4}{3}$/$\frac{4}{3}$&0/0&$-\frac{8}{3}$/$\frac{4}{3}$&0/0\\
\bottomrule[0.8pt]
\end{tabular}
\end{table}

Then the effective potentials $V_{[H_1H_2]^I_J}$ in Eq. (\ref{VHH})
for the $\bar{D}\bar{D}_s$, $\bar{D}\bar{D}_s^*$
($\bar{D}_s\bar{D}_s^*$), $\bar{D}_s^*\bar{D}_s^*$ can be
collectively written as
\begin{eqnarray}
V_{[H_1H_2]_J^I}=\tilde{g}_s\left(\mathcal{O}_1+\mathcal{O}_3\right)+\tilde{g}_a\left(\mathcal{O}_4+\mathcal{O}_6\right).\label{VMM}
\end{eqnarray}

With the parameters solved in Ref. \cite{Chen:2021cfl}, we
substitute $V_{[H_1H_2]_J^I}$ into Eq. (\ref{LSE}) to search for
possible bound states and present them in Table \ref{MMBE}. We do
not find molecular candidates in the
$\bar{D}^{(*)}_s\bar{D}^{(*)}_s$ systems since their corresponding
effective potentials are all repulsive (positive), which can be
easily checked from Table \ref{MMME} and Eq. (\ref{VMM}).

\begin{table}[!htbp]
\renewcommand\arraystretch{1.5}
\setlength\tabcolsep{0.9pt} \caption{The predicted masses and
binding energies (BE) for the considered di-meson ($[H_1H_2]_J^{I}$)
systems. Their bottomed partners are also presented. We adopt the
isospin averaged masses for the single-charm (bottom) hadrons
\cite{ParticleDataGroup:2020ssz}. The values are all in units of
MeV. \label{MMBE}}
\begin{tabular}{ccccccccccccccccccc}
\toprule[0.8pt]
System&Mass&BE&System&Mass&BE\\
\hline
$[\bar{D}\bar{D}_s]_0^{\frac{1}{2}A}$                     &3834.6  &$-1.6$ &$[BB_s]_0^{\frac{1}{2}A}$&10627.0&$-19.3$\\
$[\bar{D}\bar{D}_s^*(\bar{D}_s\bar{D}^*)]^{\frac{1}{2}A}_1$&3976.4  &$-2.1$ &$[BB_s^*(B_sB^*)]_1^{\frac{1}{2}A}$&10673.8&$-19.4$\\
$[\bar{D}^*\bar{D}_s^*]_0^{\frac{1}{2}A}$                 &4106.1  &$-14.7$&$[B^*B^*_s]_0^{\frac{1}{2}A}$&10703.7&$-36.4$ \\
$[\bar{D}^*\bar{D}_s^*]_1^{\frac{1}{2}A}$                 &4112.7  &$-8.0$&$[B^*B^*_s]_0^{\frac{1}{2}A}$&10712.2&$-27.9$\\
$[\bar{D}^*\bar{D}_s^*]_2^{\frac{1}{2}A}$                 &4120.8  &$-0.0$ &$[B^*B^*_s]_0^{\frac{1}{2}A}$&10728.7&$-11.4$\\
\bottomrule[0.8pt]
\end{tabular}
\end{table}

As presented in Table \ref{MMBE}, we obtain five possible molecular
states in the $\bar{D}^{(*)}\bar{D}_s^{(*)}$ systems. Note that for
the $[\bar{D}^*\bar{D}_s^*]_2^{\frac{1}{2}A}$ system, the
corresponding attractive force is just enough to form a bound state
with a very tiny binding energy. Besides, a common feature of these
five molecular candidates is that their flavor wave functions are
all anti-symmetric.

Since the expressions of the flavor wave functions constructed from
Eq. (\ref{MMBBFSU2}) are identical to that in the SU(3)
$\bar{\bm{3}}$ and $\bm{6}$ flavor multiplets, the
$\bar{D}_{(s)}^{(*)}\bar{D}_{(s)}^{(*)}$ systems provide us a
perfect platform to discuss the possible flavor-spin symmetry of
their interactions. In Table \ref{FSSMM}, we collect the total
effective potentials for all the
$\bar{D}_{(s)}^{(*)}\bar{D}_{(s)}^{(*)}$ systems obtained from Ref.
\cite{Chen:2021cfl} and this work.

From Table \ref{FSSMM}, we find that the flavor-spin symmetry
manifests itself clearly in the
$\bar{D}_{(s)}^{(*)}\bar{D}_{(s)}^{(*)}$ systems. In flavor space,
the anti-symmetric $\bar{D}^{(*)}\bar{D}_{(s)}^{(*)}$ systems belong
to the SU(3) $\bar{\bm{3}}$ representation, while the symmetric
$\bar{D}^{(*)}_{(s)}\bar{D}_{(s)}^{(*)}$ systems belong to the
$\bm{6}$ representation. As presented in Table \ref{FSSMM}, the
doubly charmed di-meson system with the same spin structure
$|l_1,S_1;l_2,S_2;j\rangle$ and belong to the same SU(3) multiplet
have the same effective potential.

\begin{table}
\caption{The total effective potentials $V_{[H_1H_2]_J^I}$ for all
the $\bar{D}^{(*)}_{(s)}\bar{D}^{(*)}_{(s)}$ di-meson
($[H_1H_2]_J^{I}$) systems obtained from Ref. \cite{Chen:2021cfl}
and this work. The corresponding flavor wave functions belong to
$\bar{\bm{3}}$ or $\bm{6}$ representation and have spin assignment
$|l_1,S_1;l_2,S_2;J\rangle$. The states in boldface are forbidden
due to the selection rule. \label{FSSMM}}
\begin{tabular}{ccc}
\toprule[0.8pt]
&$\bar{\bm{3}}$&$V_{[H_1H_2]^I_J}$\\
\hline
$|\frac{1}{2},0;\frac{1}{2},0;0\rangle$&\boldsymbol{$[\bar{D}\bar{D}]_0^{0}$},$[\bar{D}\bar{D}_s]_{0}^{\frac{1}{2}A}$&$-\frac{8}{3}\tilde{g}_s$\\
$|\frac{1}{2},0;\frac{1}{2},1;1\rangle$&$[\bar{D}\bar{D}^*]_1^{0}$,$[\bar{D}\bar{D}^*_s]_1^{\frac{1}{2}A}$&$-\frac{8}{3}\tilde{g}_s$\\
$|\frac{1}{2},1;\frac{1}{2},1;0\rangle$&\boldsymbol{$[\bar{D}^*\bar{D}^*]_0^0$},$[\bar{D}^*\bar{D}^*_s]_0^{\frac{1}{2}A}$&$-\frac{8}{3}\tilde{g}_s+\frac{16}{3}\tilde{g}_a$\\
$|\frac{1}{2},1;\frac{1}{2},1;1\rangle$&$[\bar{D}^*\bar{D}^*]_1^0$,$[\bar{D}^*\bar{D}^*_s]_1^{\frac{1}{2}A}$&$-\frac{8}{3}\tilde{g}_s+\frac{8}{3}\tilde{g}_a$\\
$|\frac{1}{2},1;\frac{1}{2},1;2\rangle$&\boldsymbol{$[\bar{D}^*\bar{D}^*]_2^0$},$[\bar{D}^*\bar{D}^*_s]_2^{\frac{1}{2}A}$&$-\frac{8}{3}\tilde{g}_s-\frac{8}{3}\tilde{g}_a$\\
\hline
&$\bm{6}$&$V_{[H_1H_2]^I_J}$\\
\hline
$|\frac{1}{2},0;\frac{1}{2},0;0\rangle$&$[\bar{D}\bar{D}]_0^{1}$,$[\bar{D}\bar{D}_s]_0^{\frac{1}{2}S}$,$[\bar{D}_s\bar{D}_s]_0^0$&$\frac{4}{3}\tilde{g}_s$\\
$|\frac{1}{2},0;\frac{1}{2},1;1\rangle$&$[\bar{D}\bar{D}^*]_1^{1}$,$[\bar{D}\bar{D}^*_s]_1^{\frac{1}{2}S}$,$[\bar{D}_s\bar{D}^*_s]_1^0$&$\frac{4}{3}\tilde{g}_s$\\
$|\frac{1}{2},1;\frac{1}{2},1;0\rangle$&$[\bar{D}^*\bar{D}^*]_0^{1}$,$[\bar{D}^*\bar{D}^*_s]_0^{\frac{1}{2}S}$,$[\bar{D}^*_s\bar{D}^*_s]_0^0$&$\frac{4}{3}\tilde{g}_s-\frac{8}{3}\tilde{g}_a$\\
$|\frac{1}{2},1;\frac{1}{2},1;1\rangle$&\boldsymbol{$[\bar{D}^*\bar{D}^*]_1^{1}$},$[\bar{D}^*\bar{D}^*_s]_1^{\frac{1}{2}S}$,\boldsymbol{$[\bar{D}^*_s\bar{D}^*_s]_1^0$}&$\frac{4}{3}\tilde{g}_s-\frac{4}{3}\tilde{g}_a$\\
$|\frac{1}{2},1;\frac{1}{2},1;2\rangle$&$[\bar{D}^*\bar{D}^*]_2^{1}$,$[\bar{D}^*\bar{D}^*_s]_2^{\frac{1}{2}S}$,$[\bar{D}^*_s\bar{D}^*_s]_2^0$&$\frac{4}{3}\tilde{g}_s+\frac{4}{3}\tilde{g}_a$\\
\bottomrule[0.8pt]
\end{tabular}
\end{table}

In our previous analysis, there exist only two molecular states in
the $\bar{D}^{(*)}\bar{D}^{(*)}$ systems, i.e., the
$[\bar{D}\bar{D}^*]_1^0$ and $[\bar{D}^*\bar{D}^*]_1^0$ states.
Correspondingly, in this work, we obtain their strange partners the
$[\bar{D}\bar{D}_s^*]_1^{\frac{1}{2}A}$ and
$[\bar{D}^*\bar{D}_s^*]^{\frac{1}{2}A}_{1}$ states, respectively. In
addition, the $\bar{D}$ ($\bar{D}^*$) and $\bar{D}_s$
($\bar{D}_s^*$) are no longer identical particles. Thus, in the
$\bar{D}^{(*)}\bar{D}_s^{(*)}$ systems, there exist three more
molecular candidates, i.e., the
$[\bar{D}\bar{D}_s]_0^{\frac{1}{2}A}$,
$[\bar{D}^*\bar{D}_{s}^*]_0^{\frac{1}{2}A}$, and
$[\bar{D}^*\bar{D}_{s}^*]_2^{\frac{1}{2}A}$. The last state may not
exist due to its tiny binding energy and theoretical uncertainties.

In our previous work, we propose the isospin criterion to explain
why the experimentally observed $T_{cc}^+$, $P_c$, and $P_{cs}$
molecular candidates prefer the lowest isospin numbers. As presented
in the above discussion of the
$\bar{D}_{(s)}^{(*)}\bar{D}_{(s)}^{(*)}$ systems, this isospin
criterion can be extended to the $I/U/V$ criterion. When the
considered $\bar{D}_{(s)}^{(*)}\bar{D}_{(s)}^{(*)}$ systems have the
lowest $I/U/V$ spins, they belong to the SU(3) $\bar{\bm{3}}$
multiplet and have the strongest attractive force from the exchange
of the light mesons.

Since the $D^+$ and $D_s^+$ is stable against the strong or
electromagnetic decay, the predicted $[DD_s]_0^{\frac{1}{2}A}$ is
expected to be a stable state. Besides, the $D^{*+}$ and $D_s^{*+}$
have a very narrow width, similar to the observed $T_{cc}^+$ state,
the predicted $[DD_s^{*}(D_sD^*)]_{1}^{\frac{1}{2}A}$ and
$[D^*D_s^*]^{\frac{1}{2}}$ should also be narrow in a loosely bound
molecular picture.

In our framework, we neglect the corrections from the heavy quarks.
The total effective potentials are all from the interactions of
their light d.o.f. Thus, the corresponding $T_{bbs}$ states have the
same effective potentials. In Table \ref{MMBE}, we also present the
binding energies of the molecular candidates in the
$B^{(*)}B_s^{(*)}$ systems. As can be seen from Table \ref{MMBE},
the more non-relativistic bottomed di-meson systems are more tightly
bound.

\subsection{Baryon-meson systems}

For the baryon-meson systems considered in this work, we present
their corresponding matrix elements in Table \ref{BMME}. Note that
the $B$--$M$ systems listed in Table \ref{BMME} do not allow the
exchange of the isospin triplet or doublet light mesons due to their
special flavor components. The matrix elements $\mathcal{O}_2$,
$\mathcal{O}_3$, $\mathcal{O}_5$, and $\mathcal{O}_6$ vanish. Thus,
the total effective potentials for the considered systems are
\begin{eqnarray}
V_{[BM]_{J}^{I}}=\tilde{g}_s\mathcal{O}_1+\tilde{g}_a\mathcal{O}_4.
\end{eqnarray}

\begin{table}[htbp]
\renewcommand\arraystretch{1.5}
\caption{The matrix elements of the operators $\mathcal{O}_1$ and
$\mathcal{O}_4$ for the considered heavy flavor baryon-meson systems
($[H_1H_2]_J^I$) listed in Table \ref{systems}. \label{BMME}}
\begin{tabular}{cccccc}
\toprule[0.8pt] 
System&$\mathcal{O}_1$&$\mathcal{O}_4$&System&$\mathcal{O}_1$&$\mathcal{O}_4$\\
\hline
$[\Lambda_c\bar{D}_s]_{\frac{1}{2}}^{0}$&$-\frac{4}{3}$&0 &$[\Lambda_c\bar{D}_s^*]^{0}_{\frac{1}{2}/\frac{3}{2}}$&$-\frac{4}{3}$/$-\frac{4}{3}$&0/0\\
$[\Xi_c\bar{D}_s]^{\frac{1}{2}}_{\frac{1}{2}}$&$\frac{2}{3}$&0&$[\Xi_c\bar{D}_s^*]_{\frac{1}{2}/\frac{3}{2}}^{\frac{1}{2}}$&$\frac{2}{3}$/$\frac{2}{3}$&0/0\\
$[\Sigma_c\bar{D}_s]_{\frac{1}{2}}^{1}$&$-\frac{4}{3}$&0&$[\Sigma_c\bar{D}_s^*]_{\frac{1}{2}/\frac{3}{2}}^1$&$-\frac{4}{3}$/$-\frac{4}{3}$&$\frac{16}{9}$/$-\frac{8}{9}$\\
$[\Xi_c^\prime \bar{D}_s]_{\frac{1}{2}}^{\frac{1}{2}}$&$\frac{2}{3}$&0&$[\Xi_c^\prime \bar{D}_s^*]_{\frac{1}{2}/\frac{3}{2}}^\frac{1}{2}$&$\frac{2}{3}$/$\frac{2}{3}$&$-\frac{8}{9}$/$\frac{4}{9}$\\
$[\Sigma_c^*\bar{D}_s]^1_{\frac{3}{2}}$&$-\frac{4}{3}$&0&
$[\Sigma_c^*\bar{D}_s^*]^1_{\frac{1}{2}/\frac{3}{2}/\frac{5}{2}}$&$-\frac{4}{3}$/$-\frac{4}{3}$/$-\frac{4}{3}$&$\frac{20}{9}$/$\frac{8}{9}$/$-\frac{4}{3}$\\
$[\Xi_c^*\bar{D}_s]_{\frac{3}{2}}^{\frac{1}{2}}$&$\frac{2}{3}$&0&
$[\Xi_c^*\bar{D}_s^*]^{\frac{1}{2}}_{\frac{1}{2}/\frac{3}{2}/\frac{5}{2}}$&$\frac{2}{3}$/$\frac{2}{3}$/$\frac{2}{3}$&$-\frac{10}{9}$/$-\frac{4}{9}$/$\frac{2}{3}$\\
$[\Omega_c\bar{D}]_{\frac{1}{2}}^{\frac{1}{2}}$&$-\frac{4}{3}$&0&
$[\Omega_c\bar{D}^*]^{\frac{1}{2}}_{\frac{1}{2}/\frac{3}{2}}$&$-\frac{4}{3}$/$-\frac{4}{3}$&$\frac{16}{9}$/$-\frac{8}{9}$\\
$[\Omega_c\bar{D}_s]_{\frac{1}{2}}^0$&$\frac{8}{3}$&0&
$[\Omega_c\bar{D}_s^*]_{\frac{1}{2}/\frac{3}{2}}^0$&$\frac{8}{3}$/$\frac{8}{3}$&$-\frac{32}{9}$/$\frac{16}{9}$\\
\bottomrule[0.8pt]
\end{tabular}
\end{table}

We substitute the total effective potentials of the considered $B$--$M$
systems into Eq. (\ref{LSE}) and find that some of the charmed $B$--$M$
systems have marginal attractive interactions that are just enough
to form bound states. Namely, the results for the considered $B$--$M$
systems are quite sensitive to the adopted parameters $\tilde{g}_s$
and $\tilde{g}_a$.

In Ref. \cite{Chen:2021cfl}, we use the central values of the
experimental masses of the $P_c(4312)$, $P_{c}(4440)$, and
$P_c(4457)$ states to solve the parameters $\tilde{g}_s$,
$\tilde{g}_a$, and $\Lambda$. Note that the uncertainties from the
experimental masses should be small due to the relatively small
experimental errors \cite{LHCb:2019kea}, and the $\tilde{g}_s$
dominates the total effective potentials since we obtained
$|\tilde{g}_s|/|\tilde{g}_a|\approx 4.0$. Thus, we assume that the
determined $\tilde{g}_s$ has at most 20\% deviation and check the
possible bound states of the $B$--$M$ systems by adjusting the value of
$\tilde{g}_s$.

\begin{table*}[!htbp]
\renewcommand\arraystretch{1.5}
\caption{The predicted masses and binding energies (BE) for the
heavy flavor baryon-meson ($[H_1H_2]_J^{I}$) systems listed in Table
\ref{systems}. The three sets of results are given with the
parameters $(0.8\tilde{g}_s, \tilde{g}_a)$, $(\tilde{g}_s,
\tilde{g}_a)$, and $(1.2\tilde{g}_s, \tilde{g}_a)$, where
$\tilde{g}_s=11.739$ GeV$^{-2}$ and $\tilde{g}_a=-2.860$ GeV$^{-2}$.
We adopt the isospin averaged masses for the single-charm hadrons
\cite{ParticleDataGroup:2020ssz}. The values are all in units of
MeV. \label{BMBE}}
\begin{tabular}{ccccccccccccccccccc}
\toprule[0.8pt]
System&\multicolumn{3}{c}{Mass}&\multicolumn{3}{c}{BE}&System&\multicolumn{3}{c}{Mass}&\multicolumn{3}{c}{BE}\\
\hline
&$0.8\tilde{g}_s$&$\tilde{g}_s$&$1.2\tilde{g}_s$&$0.8\tilde{g}_s$&$\tilde{g}_s$&$1.2\tilde{g}_s$&&$0.8\tilde{g}_s$&$\tilde{g}_s$&$1.2\tilde{g}_s$&$0.8\tilde{g}_s$&$\tilde{g}_s$&$1.2\tilde{g}_s$\\
\hline $[\Lambda_c
\bar{D}_s]_{\frac{1}{2}}^0$&$-$&$-$&$-$&$-$&$-$&$-$
&$[\Lambda_b B_s]_{\frac{1}{2}}^0$&10985.0&10982.6&10979.7&$-1.5$&$-3.9$&$-6.7$\\
$[\Lambda_c\bar{D}_s^*]_{\frac{1}{2}/\frac{3}{2}}^0$&$-$&$-$&$-$&$-$&$-$&$-$
&$[\Lambda_bB_s^*]_{\frac{1}{2}/\frac{3}{2}}^0$&11033.5&11031.1&11028.2&$-1.5$&$-3.9$&$-6.8$\\
$[\Sigma_c\bar{D}_s]_{\frac{1}{2}}^1$&$-$&$-$&$-$&$-$&$-$&$-$
&$[\Sigma_bB_s]_{\frac{1}{2}}^1$&11178.4&11175.9&11173.1&$-1.6$&$-4.0$&$-6.9$\\
$[\Sigma_c\bar{D}_s^*]_{\frac{1}{2}}^1$&$-$&$-$&4565.4&$-$&$-$&$-0.4$
&$[\Sigma_bB_s^*]_{\frac{1}{2}}^1$&11222.6&11219.6&11216.4&$-5.9$&$-8.9$&$-12.1$\\
$[\Sigma_c\bar{D}_s^*]_{\frac{3}{2}}^1$&$-$&$-$&$-$&$-$&$-$&$-$
&$[\Sigma_bB_s^*]_{\frac{3}{2}}^1$&11228.2&11226.5&11223.9&$-0.3$&$-2.0$&$-4.6$\\
$[\Sigma_c^*\bar{D}_s]_{\frac{3}{2}}^1$&$-$&$-$&$-$&$-$&$-$&$-$
&$[\Sigma_b^* B_s]_{\frac{3}{2}}^1$&11197.8&11195.4&11192.5&$-1.6$&$-4.1$&$-6.9$\\
$[\Sigma_c^*\bar{D}_s^*]^1_{\frac{1}{2}}$&$-$&4630.3&4629.4&$-$&$-0.1$&$-0.9$
&$[\Sigma_b^* B_s^*]_{\frac{1}{2}}$&11240.8&11237.7&11234.5&$-7.1$&$-10.2$&$-13.5$\\
$[\Sigma_c^*\bar{D}_s^*]_{\frac{3}{2}}$&$-$&$-$&$4630.3$&$-$&$-$&$-0.0$
&$[\Sigma_b^* B_s^*]_{\frac{3}{2}}$&11244.3&11241.5&11238.4&$-3.6$&$-6.4$&$-9.5$\\
$[\Sigma_c^*\bar{D}_s^*]_{\frac{5}{2}}$&$-$&$-$&$-$&$-$&$-$&$-$
&$[\Sigma_b^* B_s^*]_{\frac{5}{2}}$&11247.9&11246.7&11244.4&$-0.0$&$-1.2$&$-3.5$\\
$[\Omega_c\bar{D}]_{\frac{1}{2}}^{\frac{1}{2}}$&$-$&$-$&$-$&$-$&$-$&$-$
&$[\Omega_bB]_{\frac{1}{2}}^{\frac{1}{2}}$&11323.9&11321.5&11318.6&$-1.7$&$-4.1$&$-7.0$\\
$[\Omega_c\bar{D}^*]_{\frac{1}{2}}^{\frac{1}{2}}$&$-$&$-$&4703.3&$-$&$-$&-0.5
&$[\Omega_bB^*]_{\frac{1}{2}}^{\frac{1}{2}}$&11364.8&11361.8&11358.6&$-6.0$&-9.0&-12.2\\
$[\Omega_c\bar{D}^*]_{\frac{3}{2}}^{\frac{1}{2}}$&$-$&$-$&$-$&$-$&$-$&$-$
&$[\Omega_bB^*]_{\frac{3}{2}}^{\frac{1}{2}}$&11370.5&11368.7&11366.1&$-0.3$&$-2.1$&$-4.7$\\
\bottomrule[0.8pt]
\end{tabular}
\end{table*}

We present the results of charmed and bottomed $B$--$M$ systems that
have negative (attractive) total effective potentials in Table
\ref{BMBE}. Here, the three sets of results are obtained with the
parameters $(0.8\tilde{g}_s, \tilde{g}_a)$, $(\tilde{g}_s,
\tilde{g}_a)$, and $(1.2\tilde{g}_s, \tilde{g}_a)$. As shown in
Table \ref{BMBE}, in the charmed sector, when we adopt the
parameters $(\tilde{g}_s,\tilde{g}_a)$, we only find one molecular
state $[\Sigma_c^*\bar{D}_s^*]_{\frac{1}{2}}^1$ with a very tiny
binding energy $-0.1$ MeV. As presented in Table \ref{BMME}, the
spin-isospin related interaction ($\mathcal{O}_4$) provides an extra
attractive force in this system. This tiny interaction together with
the attraction from the exchange of isospin singlet field is just
enough to form a bound state. When we enlarge the coupling parameter
to $1.2\tilde{g}_s$, the marginal attractive forces are also
magnified. Thus we obtain a few more molecular candidates. If we
slightly decrease the coupling parameter into $0.8\tilde{g}_s$, we
find no bound state solutions..

The significant role of the $b$ quark in stabilizing the
double-bottom $B$--$M$ bound states are demonstrated in Table
\ref{BMBE}. When we adopt the lower limit of our parameter set
$(0.8\tilde{g}_s,\tilde{g}_a)$, the considered double-bottom $B$--$M$
systems all have bound state solutions. Thus, a tiny attractive
force together with a large reduced mass may lead to a bottomed
$B$--$M$ bound state. Therefore, it is not uncommon to find bound state
solutions in the doubly bottomed $B$--$M$ systems.

After studying the systems listed in Table \ref{BMME}, we go back to
the previously discussed $\Lambda_c\bar{D}^{(*)}$,
$\Sigma^{(*)}_c\bar{D}^{(*)}$, $\Xi_c\bar{D}^{(*)}$, and
$\Xi_{c}^{\prime(*)}\bar{D}^{(*)}$ systems in Ref.
\cite{Chen:2021cfl}. We find that there exist several molecular
candidates in the $\Sigma^{(*)}_c\bar{D}^{(*)}$,
$\Xi_c\bar{D}^{(*)}$, and $\Xi_{c}^{\prime(*)}\bar{D}^{(*)}$
systems. In these systems, the exchange of the isospin triplet
mesons is allowed and its contribution dominates the whole effective
potentials.

Compared with the $B$--$M$ systems studied in this work, the
$\Sigma^{(*)}_c\bar{D}^{(*)}$, $\Xi_c\bar{D}^{(*)}$, and
$\Xi_{c}^{\prime(*)}\bar{D}^{(*)}$ systems are three particularly
interesting systems in the whole double-charm $B$--$M$ di-hadron
community. They not only allow the exchange of the isospin singlet
mesons but also the exchange of the isospin triplet mesons, which
may contribute enough attractions to form bound states. On the other
hand, when we construct the wave functions of the considered $B$--$M$
systems, we consider the SU(3) breaking effects. Although the
constructed flavor wave functions of the
$\Sigma^{(*)}_c\bar{D}^{(*)}$, $\Xi_c\bar{D}^{(*)}$, and
$\Xi_{c}^{\prime(*)}\bar{D}^{(*)}$ systems may not have their
corresponding SU(3) symmetry states, we can still relate the $B$--$M$
states with the same flavor and spin matrix elements by a
generalized flavor-spin symmetry.

\begin{table}[!htbp]
\renewcommand\arraystretch{1.5}
\caption{The possible sets of molecular states studied in Ref.
\cite{Chen:2021cfl} that have identical total effective potentials
extracted from their light d.o.f. \label{SU2BM}}
\begin{tabular}{cc}
\toprule[0.8pt]
Systems&$V_{[H_1H_2]_J^I}$\\
\hline
$[\Sigma_c\bar{D}]_{\frac{1}{2}}^{\frac{1}{2}}$, $[\Sigma_c^*\bar{D}]_{\frac{3}{2}}^{\frac{1}{2}}$,$[\Xi_c\bar{D}]_{\frac{1}{2}}^{0}$,&\multirow{2}{*}{$-\frac{10}{3}\tilde{g}_s$}\\
$[\Xi_c\bar{D}^*]_{\frac{1}{2},\frac{3}{2}}^0$, $[\Xi_c^\prime\bar{D}]_{\frac{1}{2}}^{0}$,$[\Xi_c^*\bar{D}]_{\frac{3}{2}}^{0}$\\
\hline
$[\Sigma_c\bar{D}^*]_{\frac{1}{2}}^{\frac{1}{2}}$,$[\Xi_c^{\prime}\bar{D}^*]^0_{\frac{1}{2}}$&$-\frac{10}{3}\tilde{g}_s+\frac{40}{9}\tilde{g}_a$\\
$[\Sigma_c\bar{D}^*]_{\frac{3}{2}}^{\frac{1}{2}}$,$[\Xi_c^{\prime}\bar{D}^*]^0_{\frac{3}{2}}$&$-\frac{10}{3}\tilde{g}_s-\frac{20}{9}\tilde{g}_a$\\
$[\Sigma^*_c\bar{D}^*]_{\frac{1}{2}}^{\frac{1}{2}}$,$[\Xi_c^*\bar{D}^*]^0_{\frac{1}{2}}$&$-\frac{10}{3}\tilde{g}_s+\frac{50}{9}\tilde{g}_a$\\
$[\Sigma^*_c\bar{D}^*]_{\frac{3}{2}}^{\frac{1}{2}}$,$[\Xi_c^*\bar{D}^*]^0_{\frac{3}{2}}$&$-\frac{10}{3}\tilde{g}_s+\frac{20}{9}\tilde{g}_a$\\
$[\Sigma^*_c\bar{D}^*]_{\frac{5}{2}}^{\frac{1}{2}}$,$[\Xi_c^*\bar{D}^*]^0_{\frac{5}{2}}$&$-\frac{10}{3}\tilde{g}_s-\frac{10}{3}\tilde{g}_a$\\
\bottomrule[0.8pt]
\end{tabular}
\end{table}

To further emphasize this point, we collect the molecular candidates
in the $\Sigma^{(*)}_c\bar{D}^{(*)}$, $\Xi_c\bar{D}^{(*)}$, and
$\Xi_{c}^{\prime(*)}\bar{D}^{(*)}$ systems into several sets and
present them in Table \ref{SU2BM}. Each set of molecular candidates
share the same total effective potentials extracted from their light
d.o.f. For example, in Ref. \cite{Chen:2021cfl} we assign the
$P_{c}(4312)$ and $P_c(4380)$ as the
$[\Sigma_c\bar{D}]_{1/2}^{1/2}$ and
$[\Sigma_c^*D]_{3/2}^{1/2}$ states, respectively.
Their binding energies calculated from the obtained total effective
potentials listed in Table \ref{SU2BM} are $-8.9$ MeV and $-9.1$
MeV, respectively, which are in good agreement with the LHCb
results. Besides, we predict that the binding energies of the
$[\Xi_c^\prime\bar{D}^*]_{1/2}^0$ and
$[\Xi_c^\prime\bar{D}^*]_{3/2}^0$ states should be very
close to the observed $P_c(4440)^+$ and $P_{c}(4457)^+$ states if
their assignments are
$[\Sigma_c\bar{D}^*]_{1/2}^{1/2}$ and
$[\Sigma_c\bar{D}^{*}]_{3/2}^{1/2}$, respectively.
The other sets of possible molecular states that have very close
binding energies among the $\Sigma^{(*)}_c\bar{D}^{(*)}$,
$\Xi_c\bar{D}^{(*)}$, and $\Xi_{c}^{\prime(*)}\bar{D}^{(*)}$ systems
are also presented in Table \ref{SU2BM}.

\subsection{Baryon-baryon systems}

The interactions in the systems consisting of two single heavy
flavor baryons may arise from the $\mathcal{O}_1$ to $\mathcal{O}_6$
operators. Their total effective potentials can be written as
\begin{eqnarray}
V_{[H_1H_2]_J^I}=\tilde{g}_s\left(\mathcal{O}_1+\mathcal{O}_2+\mathcal{O}_3\right)+\tilde{g}_a\left(\mathcal{O}_4+\mathcal{O}_5+\mathcal{O}_6\right).\nonumber\\
\end{eqnarray}
We present the results of the $\mathcal{O}_1$-$\mathcal{O}_6$
operators for the considered di-baryon systems in Table \ref{BBME}.

\begin{table*}
\caption{The matrix elements of the operators
$\mathcal{O}_1-\mathcal{O}_6$ for the considered heavy flavor
baryon-baryon systems ($[H_1H_2]_J^I$) listed in Table
\ref{systems}. \label{BBME}}
\begin{tabular}{cccccccc}
\toprule[0.8pt]
System&$\mathcal{O}_1$&$\mathcal{O}_2$&$\mathcal{O}_3$&$\mathcal{O}_4$&$\mathcal{O}_5$&$\mathcal{O}_6$\\
\hline
$[\Lambda_c\Sigma_c^*]^{1}_{0,1}$&$\frac{4}{3}$/$\frac{4}{3}$&0/0&0/0&0/0&0/0&0/0\\
\hline
$[\Lambda_c\Xi_c]^{\frac{1}{2}S}_{0,1}$&$-\frac{2}{3}$/$-\frac{2}{3}$&0/0&2/2&0/0&0/0&0/0\\
\hline
$[\Lambda_c\Xi_c]^{\frac{1}{2}A}_{0,1}$&$-\frac{2}{3}$/$-\frac{2}{3}$&0/0&-2/-2&0/0&0/0&0/0\\
\hline
$[\Lambda_c\Xi_c^\prime]^{\frac{1}{2}S/A}_{0,1}$&$-\frac{2}{3}$/$-\frac{2}{3}$&0/0&0/0&0/0&0/0&0/0\\
\hline
$[\Lambda_c\Xi_c^*]^{\frac{1}{2}S/A}_{1,2}$&$-\frac{2}{3}$/$-\frac{2}{3}$&0/0&0/0&0/0&0/0&0/0\\
\hline
$[\Lambda_c\Omega_c]^{0S/A}_{0,1}$&$-\frac{8}{3}$/$-\frac{8}{3}$&0/0&0/0&0/0&0/0&0/0\\
\hline
$[\Sigma_c\Xi_c]^{\frac{1}{2}S/A}_{0,1}$&$-\frac{2}{3}$/$-\frac{2}{3}$&-4/-4&0/0&0/0&0/0&0/0\\
\hline
$[\Sigma_c\Xi_c]^{\frac{3}{2}S/A}_{0,1}$&$-\frac{2}{3}$/$-\frac{2}{3}$&2/2&0/0&0/0&0/0&0/0\\
\hline
$[\Sigma_c\Xi_c^{\prime}]^{\frac{1}{2}S}_{0,1}$&$-\frac{2}{3}$/$-\frac{2}{3}$&$-4/-4$&-2/-2&$\frac{8}{9}$/$-\frac{8}{27}$&$\frac{16}{3}$/$-\frac{16}{9}$&$\frac{8}{3}$/$-\frac{8}{9}$\\
\hline
$[\Sigma_c\Xi_c^{\prime}]^{\frac{1}{2}A}_{0,1}$&$-\frac{2}{3}$/$-\frac{2}{3}$&$-4/-4$&2/2&$\frac{8}{9}$/$-\frac{8}{27}$&$\frac{16}{3}$/$-\frac{16}{9}$&$-\frac{8}{3}$/$\frac{8}{9}$\\
\hline
$[\Sigma_c\Xi_c^{\prime}]^{\frac{3}{2}S}_{0,1}$&$-\frac{2}{3}$/$-\frac{2}{3}$&2/2&4/4&$\frac{8}{9}$/$-\frac{8}{27}$&$-\frac{8}{3}$/$\frac{8}{9}$&$-\frac{16}{3}$/$\frac{16}{9}$\\
\hline
$[\Sigma_c\Xi_c^{\prime}]^{\frac{3}{2}A}_{0,1}$&$-\frac{2}{3}$/$-\frac{2}{3}$&2/2&-4/-4&$\frac{8}{9}$/$-\frac{8}{27}$&$-\frac{8}{3}$/$\frac{8}{9}$&$\frac{16}{3}$/-$\frac{16}{9}$\\
\hline
$[\Sigma_c\Xi_c^{*}]^{\frac{1}{2}S}_{1,2}$&$-\frac{2}{3}$/$-\frac{2}{3}$&$-4$/$-4$&-2/-2&$\frac{20}{27}$/$-\frac{4}{9}$&$\frac{40}{9}$/$-\frac{8}{3}$&$\frac{20}{9}$/$-\frac{4}{3}$\\
\hline
$[\Sigma_c\Xi_c^{*}]^{\frac{1}{2}A}_{1,2}$&$-\frac{2}{3}$/$-\frac{2}{3}$&$-4$/$-4$&2/2&$\frac{20}{27}$/$-\frac{4}{9}$&$\frac{40}{9}$/$-\frac{8}{3}$&$-\frac{20}{9}$/$\frac{4}{3}$\\
\hline
$[\Sigma_c\Xi_c^{*}]^{\frac{3}{2}S}_{1,2}$&$-\frac{2}{3}$/$-\frac{2}{3}$&2/2&4/4&$\frac{20}{27}$/$-\frac{4}{9}$&$-\frac{20}{9}$/$\frac{4}{3}$&$-\frac{40}{9}$/$\frac{8}{3}$\\
\hline
$[\Sigma_c\Xi_c^{*}]^{\frac{3}{2}A}_{1,2}$&$-\frac{2}{3}$/$-\frac{2}{3}$&2/2&-4/-4&$\frac{20}{27}$/$-\frac{4}{9}$&$-\frac{20}{9}$/$\frac{4}{3}$&$\frac{40}{9}$/$-\frac{8}{3}$\\
\hline
$[\Sigma_c\Omega_c]^{1S/A}_{0,1}$&$-\frac{8}{3}$/$-\frac{8}{3}$&0/0&0/0&$\frac{32}{9}$/$-\frac{32}{27}$&0/0&0/0\\
\hline
$[\Sigma^*_c\Xi_c]^{\frac{1}{2}S/A}_{1,2}$&$-\frac{2}{3}$/$-\frac{2}{3}$&$-4$/$-4$&0/0&0/0&0/0&0/0\\
\hline
$[\Sigma^*_c\Xi_c]^{\frac{3}{2}S/A}_{1,2}$&$-\frac{2}{3}$/$-\frac{2}{3}$&2/2&0/0&0/0&0/0&0/0\\
\hline
$[\Sigma^*_c\Xi_c^{\prime}]^{\frac{1}{2}S}_{1,2}$&$-\frac{2}{3}$/$-\frac{2}{3}$&$-4$/$-4$&-2/-2&$\frac{20}{27}$/$-\frac{4}{9}$&$\frac{40}{9}$/$-\frac{8}{3}$&$\frac{20}{9}$/$-\frac{4}{3}$\\
\hline
$[\Sigma^*_c\Xi_c^{\prime}]^{\frac{1}{2}A}_{1,2}$&$-\frac{2}{3}$/$-\frac{2}{3}$&$-4$/$-4$&2/2&$\frac{20}{27}$/$-\frac{4}{9}$&$\frac{40}{9}$/$-\frac{8}{3}$&$-\frac{20}{9}$/$\frac{4}{3}$\\
\hline
$[\Sigma^*_c\Xi_c^{\prime}]^{\frac{3}{2}S}_{1,2}$&$-\frac{2}{3}$/$-\frac{2}{3}$&2/2&4/4&$\frac{20}{27}$/$-\frac{4}{9}$&$-\frac{20}{9}$/$\frac{4}{3}$&$-\frac{40}{9}$/$\frac{8}{3}$\\
\hline
$[\Sigma^*_c\Xi_c^{\prime}]^{\frac{3}{2}A}_{1,2}$&$-\frac{2}{3}$/$-\frac{2}{3}$&2/2&$-4$/$-4$&$\frac{20}{27}$/$-\frac{4}{9}$&$-\frac{20}{9}$/$\frac{4}{3}$&$\frac{40}{9}$/$-\frac{8}{3}$\\
\hline
$[\Sigma^*_c\Xi_c^{*}]^{\frac{1}{2}S}_{0,1,2,3}$&$-\frac{2}{3}$/$-\frac{2}{3}$/$-\frac{2}{3}$/$-\frac{2}{3}$&$-4$/$-4$/$-4$/$-4$&$-2$/$-2$/$-2$/$-2$&$\frac{10}{9}$/$\frac{22}{27}$/$\frac{2}{9}$/$-\frac{2}{3}$&$\frac{20}{3}$/$\frac{44}{9}$/$\frac{4}{3}$/$-4$&$\frac{10}{3}$/$\frac{22}{9}$/$\frac{2}{3}$/$-2$\\
\hline
$[\Sigma^*_c\Xi_c^{*}]^{\frac{1}{2}A}_{0,1,2,3}$&$-\frac{2}{3}$/$-\frac{2}{3}$/$-\frac{2}{3}$/$-\frac{2}{3}$&$-4$/$-4$/$-4$/$-4$&$2$/$2$/$2$/$2$&$\frac{10}{9}$/$\frac{22}{27}$/$\frac{2}{9}$/$-\frac{2}{3}$&$\frac{20}{3}$/$\frac{44}{9}$/$\frac{4}{3}$/$-4$&$-\frac{10}{3}$/$-\frac{22}{9}$/$-\frac{2}{3}$/$2$\\
\hline
$[\Sigma^*_c\Xi_c^{*}]^{\frac{3}{2}S}_{0,1,2,3}$&$-\frac{2}{3}$/$-\frac{2}{3}$/$-\frac{2}{3}$/$-\frac{2}{3}$&2/2/2/2&4/4/4/4&$\frac{10}{9}$/$\frac{22}{27}$/$\frac{2}{9}$/$-\frac{2}{3}$&$-\frac{10}{3}$/$-\frac{22}{9}$/$-\frac{2}{3}$/2&$-\frac{20}{3}$/$-\frac{44}{9}$/$-\frac{4}{3}$/4\\
\hline
$[\Sigma^*_c\Xi_c^{*}]^{\frac{3}{2}A}_{0,1,2,3}$&$-\frac{2}{3}$/$-\frac{2}{3}$/$-\frac{2}{3}$/$-\frac{2}{3}$&2/2/2/2&$-4$/$-4$/$-4$/$-4$&$\frac{10}{9}$/$\frac{22}{27}$/$\frac{2}{9}$/$-\frac{2}{3}$&$-\frac{10}{3}$/$-\frac{22}{9}$/$-\frac{2}{3}$/2&$\frac{20}{3}$/$\frac{44}{9}$/$\frac{4}{3}$/$-4$\\
\hline
$[\Sigma^*_c\Omega_c]^{1S/A}_{1,2}$&$-\frac{8}{3}$/$-\frac{8}{3}$&0/0&0/0&$\frac{80}{27}$/$-\frac{16}{9}$&0/0&0/0\\
\hline
$[\Xi_c\Omega_c]^{\frac{1}{2}S/A}_{1,2}$&$\frac{4}{3}$/$\frac{4}{3}$&0/0&0/0&0/0&0/0&0/0\\
\hline
$[\Xi_c^{\prime}\Omega_c]^{\frac{1}{2}S}_{0,1}$&$\frac{4}{3}$/$\frac{4}{3}$&0/0&4/4&$-\frac{16}{9}$/$\frac{16}{27}$&0/0&$-\frac{16}{3}$/$\frac{16}{9}$\\
\hline
$[\Xi_c^{\prime}\Omega_c]^{\frac{1}{2}A}_{0,1}$&$\frac{4}{3}$/$\frac{4}{3}$&0/0&-4/-4&$-\frac{16}{9}$/$\frac{16}{27}$&0/0&$\frac{16}{3}$/$-\frac{16}{9}$\\
\hline
$[\Xi_c^*\Omega_c]^{\frac{1}{2}S}_{1,2}$&$\frac{4}{3}$/$\frac{4}{3}$&0/0&4/4&$-\frac{40}{27}$/$\frac{8}{9}$&0/0&$-\frac{40}{9}$/$\frac{8}{3}$\\
\hline
$[\Xi_c^*\Omega_c]^{\frac{1}{2}A}_{1,2}$&$\frac{4}{3}$/$\frac{4}{3}$&0/0&-4/-4&$-\frac{40}{27}$/$\frac{8}{9}$&0/0&$\frac{40}{9}$/$-\frac{8}{3}$\\
\hline
$[\Omega_c\Omega_c]^0_{0}$&$\frac{16}{3}$&0&0&$-\frac{64}{9}$&0&0\\
\bottomrule[0.8pt]
\end{tabular}
\end{table*}

Note that we adopt Eq. (\ref{MMBBFSU2}) to construct the wave
functions of the considered $B$--$B$ systems. The symmetric property of
the $u$ and $d$ quarks are described by the SU(2) CG coefficients,
and we further introduce a symmetric factor to partly include the
symmetric property from the $s$ quark. As presented in Table
\ref{BBME}, the signs of the matrix elements for the $\mathcal{O}_3$
or $\mathcal{O}_6$ operator which are related to the exchange of the
strange isospin doublet fields are opposite for each $S/A$ di-baryon
systems. For the systems that do not have the contributions from the
exchange of the isospin doublet fields, we list the results of their
symmetric/antisymmetric ($S/A$) systems together since their
corresponding $\mathcal{O}_1$, $\mathcal{O}_2$, $\mathcal{O}_4$, and
$\mathcal{O}_5$ matrix elements have the identical signs.

As presented in Table \ref{BBME}, the $\Lambda_c$ and $\Xi_c$
baryons can form four different states, i.e., the
$[\Lambda_c\Xi_c]_0^{\frac{1}{2}S}$,
$[\Lambda\Xi_c]_1^{\frac{1}{2}S}$,
$[\Lambda_c\Xi_c]_0^{\frac{1}{2}A}$ and
$[\Lambda_c\Xi_c]_1^{\frac{1}{2}A}$. The matrix elements of the
$\mathcal{O}_2$ and $\mathcal{O}_5$ operators are 0 since the
exchange of the isospin triplet fields is forbidden. Moreover,
although the matrix elements of the $\mathcal{O}_1$ and
$\mathcal{O}_3$ operators are nonzero, the contributions of their
spin-related operators $\mathcal{O}_4$ and $\mathcal{O}_6$ vanish
since the $\Lambda_c$ and $\Xi_c$ contain the spin-0 light
di-quarks. In our framework, the total effective potentials are
extracted from their light d.o.f. Thus, the
$[\Lambda_c\Xi_c]^{\frac{1}{2}S/A}$ systems with $J=0$ and $1$ are
degenerate and we present them together in Table \ref{BBME}. The
inclusion of the interactions from their heavy d.o.f may distinguish
their total spins. The matrix elements of the
$\mathcal{O}_1$-$\mathcal{O}_6$ operators for the other considered
heavy flavor di-baryon systems can be understood in a similar way.

\begin{table*}[!htbp]
\renewcommand\arraystretch{1.5}
\caption{The predicted masses and binding energies (BE) for the
charmed di-baryon ($[H_1H_2]_J^{I}$) systems listed in Table
\ref{systems}. We adopt the isospin averaged masses for the
single-charm hadrons \cite{ParticleDataGroup:2020ssz}. The values
are all in units of MeV. \label{BBBE}}
\setlength{\tabcolsep}{0.4mm}{
\begin{tabular}{ccccccccccccccccccc}
\toprule[0.8pt]
System&$[\Lambda_c\Xi_c]_{0,1}^{\frac{1}{2}A}$&$[\Lambda_c\Omega_c]_{0,1}^{0S/A}$&
$[\Sigma_c\Xi_c]_{0,1}^{0S/A}$
&$[\Sigma_c\Xi_c^\prime]_{0}^{\frac{1}{2}S}$
&$[\Sigma_c\Xi_c^\prime]_{1}^{\frac{1}{2}S}$
&$[\Sigma_c\Xi_c^\prime]_{0}^{\frac{1}{2}A}$
&$[\Sigma_c\Xi_c^\prime]_{1}^{\frac{1}{2}A}$ &$[\Sigma_c\Xi_c^\prime]_0^{\frac{3}{2}A}$ &$[\Sigma_c\Xi_c^\prime]_1^{\frac{3}{2}A}$  &$[\Sigma_c\Xi_c^*]_1^{\frac{1}{2}S}$ &$[\Sigma_c\Xi_c^*]_2^{\frac{1}{2}S}$\\
Mass&4750.8  &4975.9  &4894.9   &4949.8   &4987.6   &5017.2   &5028.7  &5017.2   &5028.7  &5021.3   &5059.0\\
BE  &$-5.1$  &$-5.8$  &$-28.0$  &$-82.5$  &$-44.8$  &$-15.2$
&$-3.6$  &$-15.2$  &-3.6  &$-78.2$
&$-40.5$\\
\hline
System&$[\Sigma_c\Xi_c^*]_1^{\frac{1}{2}A}$&$[\Sigma_c\Xi_c^*]_2^{\frac{1}{2}A}$
&$[\Sigma_c\Xi_c^*]_1^{\frac{3}{2}A}$&$[\Sigma_c\Xi_c^*]_2^{\frac{3}{2}A}$&$[\Sigma_c\Omega_c]_0^{1S/A}$
&$[\Sigma_c\Omega_c]_1^{1S/A}$
&$[\Sigma_c^*\Xi_c]_{1,2}^{\frac{1}{2}S/A}$&$[\Sigma_c^*\Xi_c^\prime]_1^{\frac{1}{2}S}$
&$[\Sigma_c^*\Xi_c^\prime]_2^{\frac{1}{2}S}$&$[\Sigma_c^*\Xi_c^\prime]_{1}^{\frac{1}{2}A}$
&$[\Sigma_c^*\Xi_c^\prime]_{2}^{\frac{1}{2}A}$\\
Mass  &5085.7   &5096.8  &5085.7  &5096.8 &5133.1  &5144.8  &4959.2  &5018.7  &5056.4  &5083.1  &5094.2\\
BE    &$-13.8$  &$-2.7$  &$-13.8$ &$-2.7$ &$-15.7$ &$-3.9$  &-28.4
&$-78.2  $&-40.6  &$-13.9$
&$-2.7$\\
\hline
System&$[\Sigma_c^*\Xi_c^\prime]_{1}^{\frac{3}{2}A}$&$[\Sigma_c^*\Xi_c^\prime]_2^{\frac{3}{2}A}$
&$[\Sigma_c^*\Xi_c^*]_0^{\frac{1}{2}S}$&$[\Sigma_c^*\Xi_c^*]_1^{\frac{1}{2}S}$&$[\Sigma_c^*\Xi_c^*]_2^{\frac{1}{2}S}$
&$[\Sigma_c^*\Xi_c^*]_3^{\frac{1}{2}S}$
&$[\Sigma_c^*\Xi_c^*]_0^{\frac{1}{2}A}$&$[\Sigma_c^*\Xi_c^*]_1^{\frac{1}{2}A}$&$[\Sigma_c^*\Xi^*_c]_2^{\frac{1}{2}A}$
&$[\Sigma_c^*\Xi_c^*]_3^{\frac{1}{2}A}$&$[\Sigma_c^*\Xi^*_c]_0^{\frac{3}{2}A}$\\
Mass&5083.1  &5094.2  &5073.5  &5083.1  &5102.1  &5130.0 &5145.8  &5149.1  &5155.4  &5162.7  &5145.8\\
BE  &$-13.9$ &$-2.7$  &$-90.6$ &-81.0   &$-62.0$ &-34.1  &$-18.3$
&$-15.0$ &$-8.7$  &$-1.4$
&$-18.3$\\
\hline
System&$[\Sigma_c^*\Xi_c^*]_1^{\frac{3}{2}A}$&$[\Sigma_c^*\Xi_c^*]_2^{\frac{3}{2}A}$&$[\Sigma_c^*\Xi_c^*]_3^{\frac{3}{2}A}$&$[\Sigma_c^*\Omega_c]_1^{1S/A}$&$[\Sigma_c^*\Omega_c]_2^{1S/A}$
&$[\Xi_c^\prime\Omega_c]_0^{\frac{1}{2}A}$& $[\Xi_c^\prime\Omega_c]_1^{\frac{1}{2}A}$ &$[\Xi_c^*\Omega_c]_1^{\frac{1}{2}A}$ &$[\Xi_c^*\Omega_c]_{2}^{\frac{1}{2}A}$\\
Mass&5149.1  &5155.4  &5162.7  &5199.0  &5210.3  &5257.7  &5269.7  &5326.2  &5337.8\\
BE  &$-15.0$ &$-8.7$  &$-1.4$  &$-14.4$ &$-2.9$  &$-16.3$ &$-4.3$  &$-15.0$ &$-3.4$\\
\hline
System&$[\Lambda_c\Xi_c]_{0,1}^{\frac{1}{2}A}$&$[\Lambda_c\Omega_c]_{0,1}^{0S/A}$&
$[\Sigma_c\Xi_c]_{0,1}^{0S/A}$
&$[\Sigma_c\Xi_c^\prime]_{0}^{\frac{1}{2}S}$
&$[\Sigma_c\Xi_c^\prime]_{1}^{\frac{1}{2}S}$
&$[\Sigma_c\Xi_c^\prime]_{0}^{\frac{1}{2}A}$
&$[\Sigma_c\Xi_c^\prime]_{1}^{\frac{1}{2}A}$ &$[\Sigma_c\Xi_c^\prime]_0^{\frac{3}{2}A}$ &$[\Sigma_c\Xi_c^\prime]_1^{\frac{3}{2}A}$  &$[\Sigma_c\Xi_c^*]_1^{\frac{1}{2}S}$ &$[\Sigma_c\Xi_c^*]_2^{\frac{1}{2}S}$\\
Mass &11396.3  &11645.1  &11563.2  &11644.9  &11683.9  &11716.1  &11731.1  &11716.1  &11731.1  &11670.1  &11709.0\\
BE   &$-20.3$  &$-20.6$  &$-46.9$  &$-103.2$  &$-64.3$  &$-32.1$  &$-17.0$  &$-32.1$  &$-17.0$  &$-98.3$  &-59.4\\
\hline
System&$[\Sigma_b\Xi_b^*]_1^{\frac{1}{2}A}$&$[\Sigma_b\Xi_b^*]_2^{\frac{1}{2}A}$
&$[\Sigma_b\Xi_b^*]_1^{\frac{3}{2}A}$&$[\Sigma_b\Xi_b^*]_2^{\frac{3}{2}A}$&$[\Sigma_b\Omega_b]_0^{1S/A}$
&$[\Sigma_b\Omega_b]_1^{1S/A}$
&$[\Sigma_b^*\Xi_b]_{1,2}^{\frac{1}{2}S/A}$&$[\Sigma_b^*\Xi_b^\prime]_1^{\frac{1}{2}S}$
&$[\Sigma_b^*\Xi_b^\prime]_2^{\frac{1}{2}S}$&$[\Sigma_b^*\Xi_b^\prime]_{1}^{\frac{1}{2}A}$
&$[\Sigma_b^*\Xi_b^\prime]_{2}^{\frac{1}{2}A}$\\
Mass  &11738.3  &11753.2  &11738.3  &11753.2  &11827.0  &11842.1  &11582.6  &11669.2  &11708.1  &11737.4  &11752.3\\
BE    &$-30.2$  &$-15.2$  &$-30.2$  &$-15.2$  &$-32.2$  &$-17.1$
&$-47.0$  &-98.3    &$-59.4$
&$-30.2$  &$-15.2$\\
\hline
System&$[\Sigma_b^*\Xi_b^\prime]_{1}^{\frac{3}{2}A}$&$[\Sigma_b^*\Xi_b^\prime]_2^{\frac{3}{2}A}$
&$[\Sigma_b^*\Xi_b^*]_0^{\frac{1}{2}S}$&$[\Sigma_b^*\Xi_b^*]_1^{\frac{1}{2}S}$&$[\Sigma_b^*\Xi_b^*]_2^{\frac{1}{2}S}$
&$[\Sigma_b^*\Xi_b^*]_3^{\frac{1}{2}S}$
&$[\Sigma_b^*\Xi_b^*]_0^{\frac{1}{2}A}$&$[\Sigma_b^*\Xi_b^*]_1^{\frac{1}{2}A}$&$[\Sigma_b^*\Xi^*_b]_2^{\frac{1}{2}A}$
&$[\Sigma_b^*\Xi_b^*]_3^{\frac{1}{2}A}$&$[\Sigma_b^*\Xi^*_b]_0^{\frac{3}{2}A}$\\
Mass &11737.4  &11752.3  &11677.3  &11687.1  &11706.5  &11735.7  &11752.9  &11756.7  &11764.3  &11775.3  &11752.9\\
BE   &$-30.2$  &$-15.2$  &$-110.6$ &$-100.8$ &$-81.3$  &$-52.2$
&$-35.0$  &$-31.2$  &$-23.6$
&$-12.5$  &$-35.0$\\
\hline
System&$[\Sigma_b^*\Xi_b^*]_1^{\frac{3}{2}A}$&$[\Sigma_b^*\Xi_b^*]_2^{\frac{3}{2}A}$&$[\Sigma_b^*\Xi_b^*]_3^{\frac{3}{2}A}$
&$[\Sigma_b^*\Omega_b]_1^{1S/A}$&$[\Sigma_b^*\Omega_b]_2^{1S/A}$
&$[\Xi_b^\prime\Omega_b]_0^{\frac{1}{2}A}$& $[\Xi_b^\prime\Omega_b]_1^{\frac{1}{2}A}$ &$[\Xi_b^*\Omega_b]_1^{\frac{1}{2}A}$ &$[\Xi_b^*\Omega_b]_{2}^{\frac{1}{2}A}$\\
Mass&11756.7  &11764.3  &11775.3  &11848.3  &11863.3  &11948.8  &11963.8  &11971.0  &11986.0\\
BE  &$-31.2$  &$-23.6$  &$-12.5$  &$-30.3$  &$-15.3$  &$-32.3$  &$-17.3$  &$-30.5$  &$-15.5$\\
\bottomrule[0.8pt]
\end{tabular}
}
\end{table*}

With the obtained effective potentials, we further check whether
they have bound state solutions. We present the possible molecular
candidates of the considered di-baryon systems in Table \ref{BBBE}.
In our previous work, we only consider the di-baryon systems that
can only exchange the isospin singlet and isospin triplet fields. We
find that the repulsive or attractive forces from the exchanges of
isospin singlet mesons are very small, while the contributions from
the exchange of isospin triplet fields dominate the total effective
potentials. And the heavy flavor di-baryon systems with lower
isospin will have larger attractive forces from the exchanges of the
isospin triplet fields. Thus, the heavy flavor di-baryon systems
with lower isospin quantum numbers are more likely to form bound
states.

In this work, we further include the di-baryon systems that may also
exchange isospin doublet fields. As can be seen from Table
\ref{BBME}, for the di-baryon systems that have negative total
effective potentials and can exchange the isospin singlet, triplet
and doublet fields simultaneously, the contributions from the
exchanges of isospin singlet fields are still trivial. In contrast,
the contributions from the exchanges of the isospin triplet and
doublet fields are dominant. The exchanges of the isospin doublet
and triplet fields play a similar role. The contributions of these
two type of interactions compete with each other in the overall
effective potentials. Thus, the di-baryon system with a higher
isospin number is still possible to form a bound state if it
receives enough attractions from the exchanges of the isospin
doublet fields. For example, the
$[\Sigma_c\Xi_c^\prime]_{0,1}^{\frac{3}{2}A}$,
$[\Sigma_c\Xi_c^*]_{1,2}^{\frac{3}{2}A}$,
$[\Sigma_c^*\Xi_c^\prime]_{1,2}^{\frac{3}{2}A}$, and
$[\Sigma_c^*\Xi_c^*]_{0,1,2,3}^{\frac{3}{2}A}$ states listed in
Table \ref{BBBE} are all good molecular candidates.

\begin{table}[!htbp]
\renewcommand\arraystretch{1.5}
\caption{The possible sets of di-baryon molecular states from Ref.
\cite{Chen:2021cfl} and this work that have identical total
effective potentials extracted from their light d.o.f.
\label{SU2BB}} \setlength{\tabcolsep}{1.0mm}{
\begin{tabular}{ccccccccccccccccccc}
\toprule[0.8pt]
Systems&$V_{[H_1H_2]_J^I}$\\
\hline
$[\Lambda_c\Xi_c]_{0,1}^{\frac{1}{2}A}$,$[\Xi_c\Xi_c]_1^0$,$[\Xi_c\Xi_c^\prime]_{0,1}^0$,&\multirow{2}{*}{$-\frac{8}{3}\tilde{g}_s$}\\
$[\Xi_c\Xi_c^*]_{1,2}^0$,$[\Lambda_c\Omega_c]_{0,1}^{0S/A}$&\\
\hline
$[\Sigma_c\Sigma_c]_0^0$,$[\Sigma_c\Xi_c^\prime]_{0}^{\frac{1}{2}S}$&$-\frac{20}{3}\tilde{g}_s+\frac{80}{9}\tilde{g}_a$\\
\hline
$[\Sigma_c\Sigma_c]_1^1$,$[\Xi_c^\prime\Xi_c^\prime]_1^1$,$[\Sigma_c\Xi_c^\prime]_1^{\frac{1}{2}A}$,&\multirow{2}{*}{$-\frac{8}{3}\tilde{g}_s-\frac{32}{27}\tilde{g}_a$}\\
$[\Sigma_c\Xi_c^\prime]_1^{\frac{3}{2}A}$,$[\Sigma_c\Omega_c]_1^{1S/A}$,$[\Xi_c^{\prime}\Omega_c]^{\frac{1}{2}A}_1$&\\
\hline
$[\Sigma_c\Sigma_c^*]_1^0$,$[\Sigma_c\Xi_c^*]_1^{\frac{1}{2}S}$&$-\frac{20}{3}\tilde{g}_s+\frac{200}{27}\tilde{g}_a$\\
$[\Sigma_c\Sigma_c^*]_2^0$,$[\Sigma_c\Xi_c^*]_2^{\frac{1}{2}S}$&$-\frac{20}{3}\tilde{g}_s-\frac{40}{9}\tilde{g}_a$\\
\hline
$[\Sigma_c\Sigma_c^*]^{1}_1$,$[\Xi_c^\prime\Xi_c^*]^0_1$,$[\Sigma_c\Xi_c^*]_1^{\frac{1}{2}A}$,$[\Sigma_c\Xi_c^*]_1^{\frac{3}{2}A}$,&\multirow{2}{*}{$-\frac{8}{3}\tilde{g}_s+\frac{80}{27}\tilde{g}_a$}\\
$[\Sigma_c^*\Omega_c]_1^{1S/A}$,$[\Sigma_c^*\Xi_c^\prime]_1^{\frac{1}{2}A}$,$[\Sigma_c^*\Xi_c^\prime]_1^{\frac{3}{2}A}$,$[\Sigma_c^*\Xi_c^\prime]_1^{\frac{1}{2}A}$&\\
\hline $[\Sigma_c\Sigma_c^*]^{1}_2$,$[\Xi_c^\prime\Xi_c^*]^0_2$,$[\Sigma_c\Xi_c^*]_2^{\frac{1}{2}A}$,$[\Sigma_c\Xi_c^*]_2^{\frac{3}{2}A}$,&\multirow{2}{*}{$-\frac{8}{3}\tilde{g}_s-\frac{16}{9}\tilde{g}_a$}\\
$[\Sigma_c^*\Omega_c]_2^{1S/A}$,$[\Sigma_c^*\Xi_c^\prime]_2^{\frac{1}{2}A}$,$[\Sigma_c^*\Xi_c^\prime]_2^{\frac{3}{2}A}$,$[\Sigma_c^*\Xi_c^\prime]_2^{\frac{1}{2}A}$&\\
\hline
$[\Sigma_c^*\Sigma_c^*]_0^0$,$[\Sigma_c^*\Xi_c^*]_0^{\frac{1}{2}S}$&$-\frac{20}{3}\tilde{g}_s+\frac{100}{9}\tilde{g}_a$\\
$[\Sigma_c^*\Sigma_c^*]_2^0$,$[\Sigma_c^*\Xi_c^*]_2^{\frac{1}{2}S}$&$-\frac{20}{3}\tilde{g}_s+\frac{20}{9}\tilde{g}_a$\\
$[\Sigma_c^*\Sigma_c^*]_1^1$,$[\Xi_c^*\Xi_c^*]_1^0$,$[\Sigma_c^*\Xi_c^*]_1^{\frac{1}{2}A}$,$[\Sigma_c^*\Xi_c^*]_1^{\frac{3}{2}A}$&$-\frac{8}{3}\tilde{g}_s+\frac{88}{27}\tilde{g}_a$\\
$[\Sigma_c^*\Sigma_c^*]_3^1$,$[\Xi_c^*\Xi_c^*]_3^0$,$[\Sigma_c^*\Xi_c^*]_3^{\frac{1}{2}A}$,$[\Sigma_c^*\Xi_c^*]_3^{\frac{3}{2}A}$&$-\frac{8}{3}\tilde{g}_s-\frac{8}{3}\tilde{g}_a$\\
\bottomrule[0.8pt]
\end{tabular}
}
\end{table}

For the studied di-baryon systems, based on the flavor and spin
structures of their light d.o.f, we can also collect the states that
have identical $\langle\bm{\lambda}_1\cdot\bm{\lambda}_2\rangle$ and
$\langle\bm{\sigma}_1\cdot\bm{\sigma}_2\rangle$ matrix elements, and
relate them with the generalized flavor-spin symmetry. In Table
\ref{SU2BB}, we collect several sets of molecular candidates that
have the same total effective potentials obtained from Ref.
\cite{Chen:2021cfl} and this work. Each set of molecular states are
expected to have very close binding energies.

\section{Summary}\label{sec4}

In our previous work, we proposed a unified formalism to describe
the observed $T_{cc}$, $P_c$, and $P_{cs}$ states. Their
interactions are related via a quark-level Lagrangian and we obtain
a satisfactory description of their masses. We also study the
possible molecular candidates in the other systems composed of two
single heavy flavor hadrons (meson-meson, baryon-meson, and
baryon-baryon systems) that can only exchange the isospin singlet
and triplet scalar and (or) axial vector fields.

In this work, we further include the systems that can also exchange
the isospin doublet fields. The inclusion of these systems allows us
to further explore the possible symmetry properties of the
interactions in the heavy flavor meson-meson, baryon-meson, or
baryon-baryon systems. This is the main task of this work. These two
works provide a complete and general discussion of the possible
molecule community composed of two single heavy flavor di-hadron
systems.

For the ground light flavor scalar and axial-vector mesons, the
masses of the isospin doublet mesons are close to those of the
isospin singlet or triplet mesons. Thus, we neglect the difference
between the isospin doublet mesons and isospin singlet or triplet
mesons and only introduce two coupling parameters $\tilde{g}_s$ and
$\tilde{g}_a$ to describe the interactions from the exchanges of the
scalar and axial vector octet mesons, respectively. The Lagrangian
has a strict SU(3) symmetry. However, in order to distinguish the
physical states from the SU(3) flavor states, we retain the SU(2)
symmetry and take the $s$ quark as a flavor singlet to construct the
flavor wave functions of the considered di-hadron systems. For the
$M$--$M$, and $B$--$B$ systems, we partly include the symmetry properties
of the $s$ quark by constructing the $[H_1H_2]^I$ and $[H_2H_1]^I$
flavor wave functions and combine them with a symmetry factor.

For the $\bar{D}^{(*)}_{(s)}\bar{D}^{(*)}_{(s)}$ systems, the
constructed flavor wave functions are identical to the expressions
of the SU(3) flavor $\bm{\bar{3}}$ and $\bm{6}$ multiplets. The
antisymmetric $\bm{\bar{3}}$ states correspond to the
$\bar{D}^{(*)}\bar{D}_{(s)}^{(*)}$ states with the lowest $I/U/V$
spins and have enough attractive forces to form bound states. The
interactions of the states that belong to the flavor $\bar{\bm{3}}$
or $\bm{6}$ multiplet with a specific $|l_1,S_1;l_2,S_2;J\rangle$
spin structure can be related by the flavor-spin symmetry induced
from our Lagrangian. The molecular candidates in the
$\bar{D}^{(*)}\bar{D}_s^{(*)}$ systems are also discussed.

Our analysis of the $B$--$M$ systems suggests that the previously
studied $\Xi_c\bar{D}^{(*)}$, $\Xi_c^{\prime(*)}\bar{D}^{(*)}$, and
$\Sigma_c^{(*)}\bar{D}^{(*)}$ systems are three particularly
important systems. Apart from the exchanges of the isospin singlet
fields, the exchanges of the isospin triplet fields are also
allowed, which dominate the total effective potentials of the
molecular states in these systems. Among them, we use a generalized
flavor-spin symmetry to relate the states with the same
$\langle\bm{\lambda}_1\cdot\bm{\lambda}_2\rangle$ and
$\langle\bm{\sigma}_1\cdot\bm{\sigma}_2\rangle$ matrix elements and
present them in Table \ref{SU2BM}. In contrast, the $B$--$M$ systems
studied in this work can only exchange the isospin singlet fields.
Even if the corresponding contributions are attractive, the weak
attractive forces can hardly form bound states in the charmed
sector. However, the more non-relativistic nature of the bottomed
$B$--$M$ systems helps to stabilize the weakly attractive systems and
form bound states.

We further analyze the doubly heavy flavor di-baryon systems. Our
results together with the results in Ref. \cite{Chen:2021cfl}
suggest that there exist rich molecular candidates in the doubly
heavy flavor di-baryon systems. We also predict the binding energies
of the di-baryon molecular candidates, and present several sets of
molecular states that share the same flavor and spin matrix
elements. The binding energies for each sets of molecules are very
close due to the generalized flavor-spin symmetry. The LHCb
collaboration may have the potential to look for these systems in
the future.
\section*{Acknowledgments}

This project is supported by the National Natural Science Foundation
of China under Grants 11975033 and 12070131001. B. Wang is supported
by the National Natural Science Foundation
of China under Grant 12105072, the Youth Funds of Hebei Province (No. A2021201027) and the
Start-up Funds for Young Talents of Hebei University (No. 521100221021).

\end{document}